\documentclass[%
 aps,
 jmp,%
 amsmath,amssymb,
 reprint,%
]{revtex4-1}

\usepackage{graphicx}% Include figure files
\usepackage{dcolumn}% Align table columns on decimal point
\usepackage{bm}% bold math
\usepackage{xcolor}
\begin{document}

%\preprint{AIP/123-QED}

\title[]{Finite Field Methods for the Supercell Modelling of Charged Insulator-Electrolyte Interfaces}% Force line breaks with \\
%\thanks{Footnote to title of article.}

\author{Chao Zhang}
\author{Michiel Sprik}
\email{ms284@cam.ac.uk}
\affiliation{Department of Chemistry, University of Cambridge,
  Lensfield Rd, Cambridge CB2 1EW, United Kingdom%\\This line break forced with \textbackslash\textbackslash
}%

\date{\today}% It is always \today, today,

\begin{abstract}
Surfaces of ionic solids interacting with an ionic solution can build up charge by exchange of ions. The surface charge is compensated by a strip of excess charge at the border of the electrolyte forming an electric double layer. These electric double layers are very hard to model using the supercells methods of computational condensed phase science. The problem arises when the solid is an electric insulator (as most ionic solids are) permitting a finite interior electric field over the width of the slab representing the solid in the supercell. The slab acts as a capacitor. The stored charge is a deficit in the solution failing to compensate fully for the solid surface charge. Here we show how these problems can be overcome using the finite field methods developed by Stengel, Spaldin and Vanderbilt [Nat. Phys. {\bf 5}, 304, (2009)]. We also show how the capacitance of the double layer can be computed once overall electric neutrality of the double layer is restored by application of a finite macroscopic field $\mathbf{E}$ or alternatively by zero electric displacement  $\mathbf{D}$. The method is validated for a classical model of a solid-electrolyte interface using the finite temperature molecular dynamics adaptation of the constant field method presented previously  [Phys. Rev. B, 2016, 93, 144201]. Because  ions in electrolytes can diffuse across supercell boundaries, this application turns out to be a critical illustration of the multivaluedness of polarization in periodic systems. 
%
%Valid PACS numbers may be entered using the \verb+\pacs{#1}+ command.
\end{abstract}

\maketitle

%\begin{quotation}
%The ``lead paragraph'' is encapsulated with the \LaTeX\ 
%\verb+quotation+ environment and is formatted as a single paragraph before the first section heading. 
%(The \verb+quotation+ environment reverts to its usual meaning after the first sectioning command.) 
%Note that numbered references are allowed in the lead paragraph.
%
%The lead paragraph will only be found in an article being prepared for the journal \textit{Chaos}.
%\end{quotation}

\section{Introduction}

Surfaces of solids in contact with an electrolyte are often charged. The example familiar from electrochemistry is the interface between a metal electrode and ionic solution\cite{Gerischer97,Sato98,Schmickler:2010Sc}. The charge on the metal is compensated by counter charge on the solution side. The electronic charge and the neutralizing excess ionic charge occupy a finite region with net zero charge called an electrical double layer (EDL). However also insulators in contact with an electrolyte form electrical double layers\cite{Westall:1980cols,Trasatti:1996cols,Israelachvili:2011Is}. The surface charge is now of chemical origin, either due to adsorption of ions from solution or desorption from the solid. An important example of this type of charged interface is an oxide surface exchanging protons with an aqueous solution. Deprotonation of terminal hydroxide groups or adsorbed water at high pH leads to a negatively charged surface and protonation of basic surface groups at low pH to positively charged surfaces\cite{Trasatti:1996cols}. Semiconductor-electrolyte interfaces usually carry both chemical and electronic charge (now also spread out over a finite width space charge region) making semiconductor electrochemistry a notoriously difficult subject\cite{Gerischer97,Sato98}. 

Solid-electrolyte interfaces have been and continue to be a major target for modelling in physics, chemistry and biology. Charged interfaces are particularly challenging for fully atomistic modelling treating both the solid and electrolyte in microscopic detail.  The origin of the problems is that a heterogeneous system consisting of a semi-infinite solid and electrolyte is incompatible with the periodic supercells used in atomistic modelling.  Periodicity in the direction perpendicular to the interface inevitably introduces a second EDL. The result is the well known geometry of alternating slabs of solid and liquid as is exemplified  by the model system studied here (Fig.~\ref{ecs}). The question is how to deal with the long range electrostatic interactions between aligned EDL dipoles and their periodic images.   

Elimination of undesirable electrostatic interactions between dipole layers in periodic supercells is one of the most intensely studied technical subjects  in computational condensed phase science. The default method for the evaluation of electrostatic interactions in electronic structure calculation is Ewald summation, performing part or all of the calculation in reciprocal space\cite{DeLeeuw:1980rs80a,DeLeeuw:1980rs80b,Ballenegger:2014jpc}. While there are alternatives for classical force field simulation, we will restrict this brief introductory overview to schemes based on Ewald summation.

 A variety of methods has been proposed. The most radical of these imposes periodicity only in the lateral directions keeping the system finite in the direction perpendicular to the interface. This requires a two dimensional (2D) version of Ewald summation\cite{Hautman:1992mp,Kawata:2001cpl,HuZ:2014jctc}. 2D Ewald summation techniques have been improved in efficiency and accuracy over time and are a good option for classical force field based models\cite{HuZ:2014jctc}. We should note that electrolyte solutions confined between two flat metallic electrodes can also be treated by Ewald-compatible image charge methods which have the advantage that the potential is constant over the metal surface by construction\cite{Perram:1996jcp,Voth:2012jpcc,Onuki:2013jcp}. The present paper is however only concerned with insulator-electrolyte interfaces. 

For electronic structure calculation 2D-Ewald is less attractive or, in practice, hardly ever used. The preferred approach is to insert a vacuum layer in a 3D periodic supercell. The long range range interactions between periodic images are removed by Coulomb cut-off\cite{Rubio:2006prb,IsmailBeigi:2006prb,Bernholc:2008prb} or correction terms\cite{Neugebauer:1992uh,Bengtsson:1999il,Yeh:1999dm,Tuckerman:2002jcp,Bernholc:2008prb,Marzari:2008prb,Yeh:2011bb,Pasquarello:2013prl}. A popular and simple correction scheme is the so called dipole correction method, developed in parallel in physics\cite{Neugebauer:1992uh,Bengtsson:1999il} and physical chemistry~\cite{Yeh:1999dm} where it is called the Yeh-Berkowitz (YB) correction. The correction term in this method can be seen as a self interaction of the total dipole moment of the slab system.  We will return to this method later.     

In the present contribution we adopt a rather different approach. Our scheme is based on the finite field methods for periodic supercells introduced by Stengel, Spaldin and Vanderbilt (SSV)~\cite{Stengel:2009cd,Stengel:2009prb}. The SSV finite field method is a recent spin-off of the modern theory of polarization developed by Vanderbilt and Resta during the 90's~\cite{King-Smith:1993prb,Resta:1994rmp,Resta:2007ch}. The modern theory of polarization caused a revolution in theoretical and computational solid state physics making it possible, for the first time, to investigate the electric equation of state of ferroelectric systems. The initial approach was to compute the total energy for fixed values of the polarization and to determine the electric field from the derivative\cite{Nunes:1994prl,Sai:2004prb,Dieguez:2006prl}. This was subsequently changed to a scheme using directly the macroscopic electric field $\mathbf{E}$ or the electric displacement field $\mathbf{D}$  as the control variable, which has both computational and conceptual advantages\cite{Umari:2002prl,Umari:2004cpl,Stengel:2009cd,Stengel:2009prb,Stengel:2009natm,Stengel:2009prr,Stengel:2011prl,Cazorla:2012prb}.

The SSV method was applied by Vanderbilt and colleagues in their computational investigation of ferroelectric nanocapacitors\cite{Stengel:2009prb,Stengel:2009natm,Stengel:2009prr}. A capacitor was modelled by a superlattice of alternating ferroelectric and metallic crystalline materials under full 3D periodic boundary conditions (PBC). This suggests a parallel to charged insulator electrolyte interfaces. In both systems the insulator slab is polar. For a ferroelectric capacitor this is due to spontaneous bulk polarization. For our insulator (oxide) the polarization is generated by excess surface charge stabilized by the surface chemistry. For the elementary model of a charged insulator-electrolyte interface studied here this is the charge denoted by  $\sigma_0$ and $-\sigma_0$ in Fig.~\ref{ecs}.  The electrolyte, while an electronic insulator, is an ionic conductor and will also screen the polarization. The electrolyte therefore plays the role of the metal in the ferroelectric capacitor. 

The parallel to ferroelectric nanocapacitors has a rather disconcerting implication. It suggests that the insulator slab  in fact acts as a capacitor storing charge at the interface with the electrolyte. This charge would not be $\sigma_0$ but a net charge of the EDL due to incomplete compensation of $\sigma_0$.  This would also imply that the interior electric field in the insulator is finite. This a direct consequence of Maxwell's law for parallel plate capacitors\cite{Purcell:2011ca}. The calculations presented in the following confirm this expectation. This is not what one would like if the objective is to model the EDL separating a semi-infinite solid and electrolyte. Such an EDL is overall neutral with zero field in the bulk solid away from the interface. The situation is very similar for nanostructures with polar surfaces\cite{Stengel:2009prr,Stengel:2011prl,Cazorla:2012prb,Noguera:2008rpp}, which offer another instructive perspective on charged insulator-electrolyte interfaces.    

The coupling between net EDL charge and the interior field in the insulator works the other way around as well: Forcing the interior field to vanish should balance the EDL's. We will refer to this state as the point of zero net charge (ZNC). There is of course an easy way of cancelling the interior field, namely using a symmetric slab with surface charge of the same sign on both sides.  This scheme was used in earlier work on density functional theory (DFT) modelling of charged ionic solid-electrolyte interfaces\cite{Cheng:2014eh,Sulpizi:2016scirep}. The drawback is that the surface charge can only be changed by removing or inserting ions. This is how the protonation state of the TiO$_2$ slab was varied in Ref.~\citenum{Cheng:2014eh}. Every  surface proton added or removed had to be balanced by adjusting the number of ions  in the electrolyte in order to maintain overall neutrality.  This amounts to a drastic change in composition for the small system sizes used in electronic structure calculation.  

The interior field in the insulator can also be cancelled by subjecting the system to a compensating external field. This way of restoring charge balance preserves composition. The surface charge is simply transferred from side of the slab to the other. The charge distribution in electrolyte adapts by polarizing. The development of the SSV finite field methods has made it possible to apply this method to a fully periodic heterogeneous system. We can either use a constant macroscopic field $\mathbf{E}$ or constant electric displacements field $\mathbf{D}$\cite{Stengel:2009cd, Stengel:2009prb}. As we will show, $\mathbf{E}$ is finite at ZNC while $\mathbf{D}=0$. Moreover, both methods can be used to estimate the (series) capacitance of the pair of EDL's. In case of the constant $\mathbf{E}$ approach the capacitance is directly obtained form the value of $\mathbf{E}$ at ZNC. However, in order to check whether the excess charge in the electrolyte cancels the surface charge $\sigma_0$, this method requires computation of the net charge of an EDL. This is not necessary for the finite $\mathbf{D}$ electric boundary conditions. The capacitance is computed from the (linear) increment of the polarization in response to a change in $\sigma_0$, which for electronic systems is easier and more in the spirit of the modern theory of polarization.            

The field assisted method outlined above will be validated in a study of the EDL model system of Fig.~\ref{ecs} applying a classical finite temperature implementation of the SSV method. We have used this approach already in two previous publications for the calculation of the dielectric constant of liquid water\cite{Zhang:2015ms, Zhang:2016ho}. The present application is however more challenging, addressing the defining feature of the modern theory of polarization, the multivaluedness of polarization\cite{Resta:2007ch}. 

The problem is particularly acute in ionic solutions because ions can cross MD cell boundaries.  Polarization in the SSV method is defined in terms of a time integral of current, which requires that the motion of the ions is continuous. This is not an issue for the supercell of Fig.~\ref{ecs}, where the ions are prevented from leaving by the rigid walls. However they can leave in the system of Fig.~\ref{ics} which is a different supercell representation of the same periodic slab system. The electric properties computed for the same physical system shown in Figs.~\ref{ecs} and \ref{ics} must be consistent. How to achieve this consistency by taking care of the multivaluedness of polarization is a major theme of this paper and a key tool in the validation of our method for the computation of capacitance.    
 
The outline of the paper is as follows:  After introducing the setup of the two alternative supercells for the periodic charged insulator-electrolyte interface, the effect of net EDL charge is quantified by a MD calculation. Next we define a Stern-like continuum model to analyze these results. We then derive an electric equation of state for the continuum model which is used to obtain expressions for the capacitance of the EDL in the Helmholtz approximation under constant $\mathbf{E}$ and constant $\mathbf{D}$ conditions. Following that, the atomistic SSV electric Hamiltonians at constant $\mathbf{E}$ and constant $\mathbf{D}$ are reviewed together with a discussion of the manifestation of multivalued polarization in our EDL model. Finally, the EDL capacitance is computed for the atomistic model using the two finite field methods suggested by the continuum with additional comments on computational efficiency. We conclude with an outlook to future applications going beyond the elementary test system used here.

\section{Double layer supercell} \label{sec:pbc}

\subsection{Periodic boundary conditions}
\label{sec:pbcmodel}
\begin{figure} [h]
  \includegraphics[width=1.0\columnwidth]{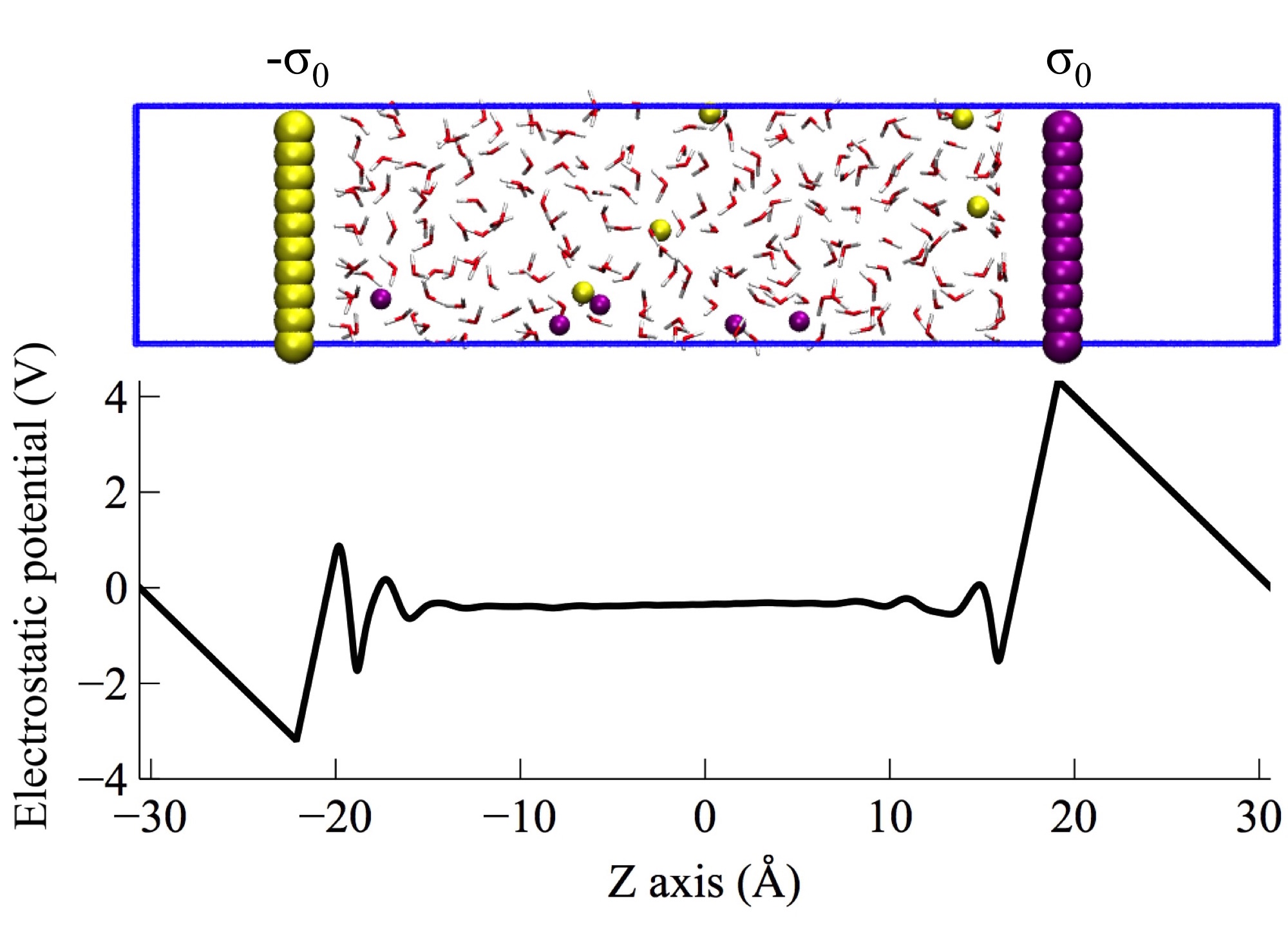}
\caption{\label{ecs} Periodic model of two complementary charged insulator-electrolyte interfaces used as the model system in this study. The charged insulator is modelled as a pair of rigid atomic walls with opposite charge separated by a vacuum region. Vacuum slab here is used as the absolutely simplest realization of an insulator. The surface charge is uniformly distributed over the 100 atoms making up a wall. Positive charges are in purple and negative charges are in yellow. The electrolyte consists of 202 water molecules, 5 Na$^+$ and 5 Cl$^-$ ions, representing a 1.4M electrolytic solution. For further details see Section \ref{sec:spcmd}. The lower panel gives the electrostatic potential profile averaged over the perpendicular $x$ and $y$ directions accumulated over 1 ns.  This configuration will be referred to as the electrolyte centred supercell (ECS).}
\end{figure}

\begin{figure} [h]
\includegraphics[width=1.0\columnwidth]{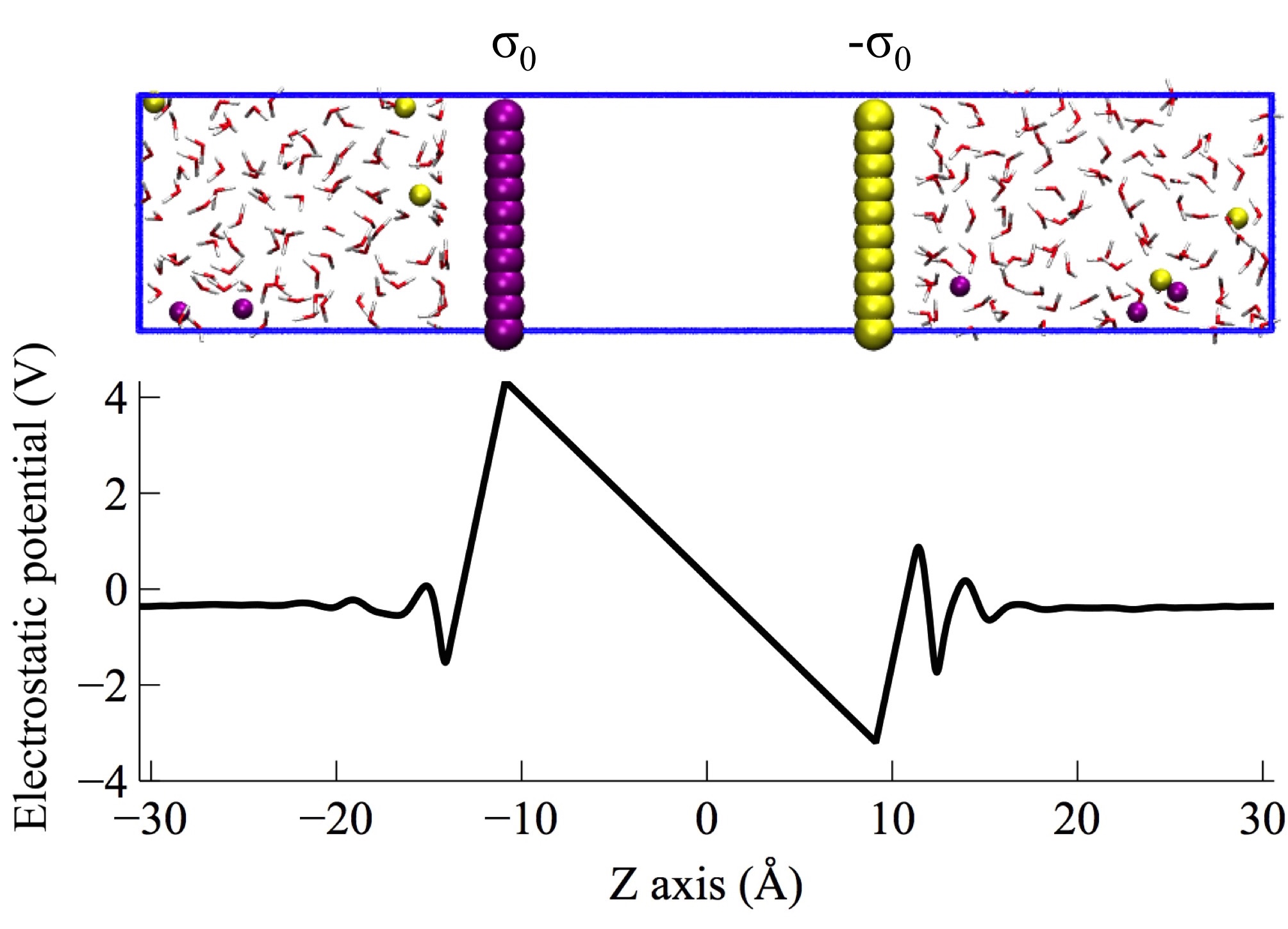}
\caption{\label{ics} Alternative view of the same charge insulator-electrolyte model system of Fig.~\ref{ecs}.  The difference is the choice of supercell. Again, vacuum slab here is used as the absolutely simplest realization of an insulator. In Fig.~\ref{ecs} the electrolyte is wholly contained in the cell with the boundaries bisecting the vacuum region. In the figure above the cell is centred on the vacuum region.  Now the electrolyte is partitioned over two parts of the cell. This configuration will be referred to as the insulator centred supercell (ICS). While the ECS and ICS models are two representations of exactly the same periodic system, the polarization is different. This problem is another manifestation of the ambiguity in the definition of polarization for extended models of ionic crystals, well-known from textbooks on solid state physics.}
\end{figure}

Periodic models of charged interfaces can be confusing, even paradoxical. To lay out the problems that need to be resolved, we thought it helpful to present the actual model system right at the beginning, discussing theory along side key results. The model is shown in Figs~\ref{ecs} and \ref{ics}. It is another example of an SPC model familiar from many studies of electrolyte-charged interfaces~\cite{Spohr:1999tm, Dimitrov:2000ti,Fedorov:2008gh,Zarzycki:2010kja,LyndenBell:2012iq, Sultan:2014hj, Parez:2014dx, Dewan:2014ig, Hocine:2016da}.   The solid surface is an atomic plane of Lennard-Jones atoms with partial point charges giving a specific surface charge density $\sigma_0$. The surface charge density is fixed and is the primary system control parameter. There are two such charged walls of opposite charge.  The space in between is filled with a SPC model of an aqueous ionic solution~\cite{Berendsen:1981spc,Joung2008}, as shown in Fig.~\ref{ecs} (1.4M of NaCl in our case, see Fig.~\ref{ecs} caption and the model description in Section \ref{sec:spcmd}).

Under PBC there are two open spaces between the charge sheets.  We have filled one with electrolyte, the other is going to be the insulator. The absolutely simplest realization of an insulator is vacuum.  The second space between the charged walls is therefore left empty. These are the white regions on the left and the right in Fig.~\ref{ecs}. When applying PBC, the  insulator slab is one piece of ``material'' as can be better seen in Fig~\ref{ics} where the centre of the supercell has moved over half the cell length to the middle of the insulator.     

The solid walls are completely rigid. The electrolyte is free to move. Defining the parameter $\sigma_0$ specifying the surface charge to be positive ($\sigma_0 \ge 0 $), the Na$^+$ ions are attracted to the wall with charge $-\sigma_0$ and will form a electric double layer (EDL) screening the surface charge. Similarly the Cl$^-$ will tend to accumulate at the opposite wall with charge $\sigma_0$. Lower panels of Figs.~\ref{ecs} and \ref{ics} show the resulting electrostatic potential profiles. These profiles have been obtained averaging the instantaneous potential over MD trajectories of 1 ns  (For consistent calculation of potential profiles generated by periodic charge densities, please refer to Refs.~\cite{Yeh:2011bb,Wirnsberger:2016wi}).  Despite the small number of ions (5 positive and 5 negative ions), the electrostatic potential in the middle of the electrolyte compartment remains flat over a length of about 20~\AA{}, as shown in the electrolyte centred supercell (ECS) Fig.~\ref{ecs}. This indicates that the net charge in this region of the electrolyte is neutral which can therefore be regarded as ``bulk'' electrolyte. The structure in the potential adjacent to the walls is a manifestation of the complex ordering of ionic charge and solvent molecules in a compact (Helmholtz) EDL formed when the ionic strength of the electrolyte is high (1.4M in our model). 

\subsection{Uncompensated double layer charge} \label{sec:edlch}
The potential profile, shown better in the insulator centred supercell (ICS) view of Fig.~\ref{ics}, brings out the main concern for supercell modelling of  double layers. Contrary to the potential in the electrolyte, the potential in the insulator (vacuum) section varies linearly in $z$, the coordinate perpendicular to the charged walls, resembling the potential across a parallel plate capacitor~\cite{Purcell:2011ca}.  The electrolyte on the left and right plays the role of the conducting electrodes with constant inner potential and zero field. Linear dependence in $z$ is the characteristic behaviour of the dielectric material in a capacitor under a finite bias voltage.

 However, the capacitor of Fig.~\ref{ics} is in short circuit. The potentials on the left and right side are the same. They must be, because of the periodic boundary conditions. There is only a single body of electrolyte as explicitly seen in the ECS view (Fig.~\ref{ecs}). The charge generating the finite field is not the result of the application of an external bias potential but is due to the incomplete charge compensation in the EDL. The number of  counter ions supplied by the electrolyte is less than required to cancel the fixed surface charge  $\sigma_0$ on the faces of the insulator. The solid state parallel that comes to mind is that of a ferroelectric crystal with electrodes attached. The electrodes are connected in short circuit. The electronic charge induced in the electrodes  stabilizes the spontaneous polarization by screening the depolarizing field. Incomplete screening, as occurs in ferroelectric nanocapacitors, can be viewed as the formation of an electrical double layer leading to deterioration of device performance~\cite{Stengel:2006kv,Stengel:2009go}.  

The implication of the  plate capacitor picture is that the charge insulator-electrolyte interface in our double layer should carry a finite charge.  This prediction can be easily verified by computing the total charge. To this end we define a charge profile $Q(z)$ according to:
\begin{equation}
\label{Q_z_md}
Q(z)  = A \int^z_{-L/2} dz' \rho(z') 
\end{equation}
where $\rho(z)$ is the laterally $(x,y)$ averaged charge density at  location $z$ in the perpendicular direction. $A$ is the area of the $x,y$ cross section. The $z$ coordinate is referred to the middle of the supercell with boundaries at $z=-L/2$ and $z=+L/2$. Integration starts at $z=-L/2$ at the boundary of the left vacuum region in Fig.~\ref{ecs}.  Therefore, $Q(z)$ is well defined here. The net charge $Q_\text{net}$ of the EDL is calculated as the value $Q(z_e)$ at any location $z_e$  in the bulk electrolyte. The choice of $z_e$ is unimportant in this case, because the field is zero in the bulk electrolyte.

\begin{figure} [h]
\includegraphics[width=0.95\columnwidth]{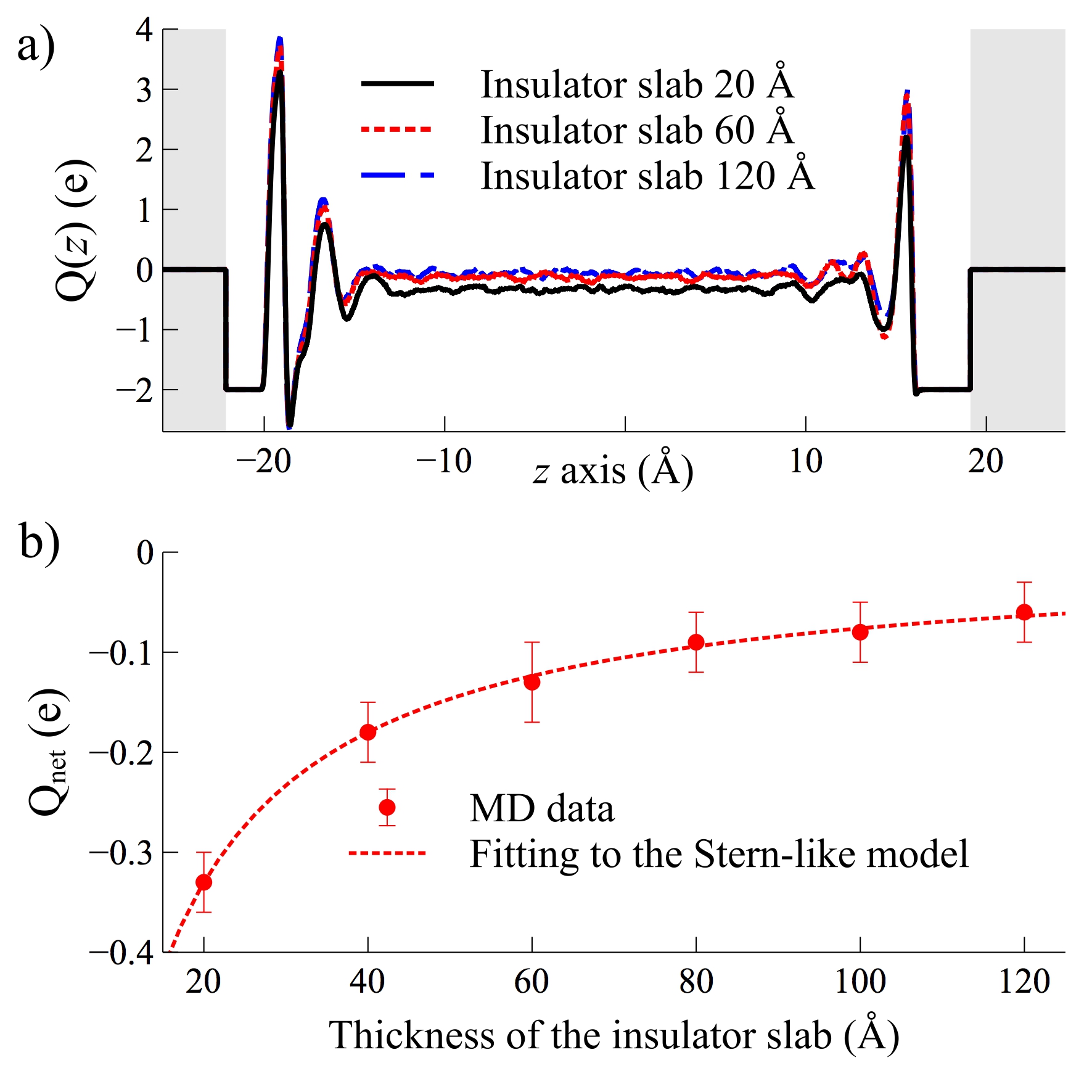}% Here is how to import EPS art
\caption{\label{Qz} a) The integrated charge profile $Q(z)$ defined in Eq.~\ref{Q_z_md} for different thickness of the insulator slab. The system is represented using an electrolyte centred supercell (Fig.~\ref{ecs}). The region of the insulator slab is shaded.; b) The net charge $Q_{\textrm{net}}$ of the left EDL as a function of the insulator slabs' thickness. $Q_{\textrm{net}}$ should be zero for an interface between a semi-infinite insulator and electrolyte. The dotted line follows the expression for the net charge $Q_{\textrm{net}}$ of the macroscopic continuum model (Eq.~\ref{Qnet_stern}, Section~\ref{sec:sternsc}).}
\end{figure}

$Q_\text{net}$ is indeed finite, as shown in Fig.~\ref{Qz}.  The fixed charge on the left insulator surface was set to $A \sigma_0 = - 2 e$ ($e$ is the unit charge). $Q_\text{net}$ is still about $-0.1 e$, 5$\%$ of the full charge, when the insulator slab is as large as 100 \AA. This is a most serious finite size error if the periodic system  is to describe the interface between a semi-infinite (macroscopic) solid and an electrolyte. The error decreases for increasing thickness of the insulator slab scaling as $1/l_\text{d}$, the thickness of the insulator slab (See Section \ref{sec:sternsc} for the proof). This means that in order to recover a faithful model of a macroscopic charged insulator-electrolyte interface, one would need a very large slab of insulator. This is already a challenge for force field-based simulations (see Fig.~\ref{Qz}) and simply not feasible for DFT-based simulations. In order to tackle this challenge, we need to understand the electrostatics and dielectrics of the charged insulator-electrolyte system with parallel EDLs under PBC. The clue is the finite field inside the dielectric, i.e. the vacuum segment of our system.

\subsection{Finite electric field in the dielectric slab} \label{sec:edielec}
 The potential of a plate capacitor is determined by the net surface charge density $q$ at the electrode-dielectric interface\cite{Purcell:2011ca}.  $q$ is the sum of the surface charge density on a metal electrode and the polarization surface charge density of the dielectric. Similarly the charge profile $Q(z)$ of  Eq.~\ref{Q_z_md} is the sum of the (fixed) charge on the insulator slab, the polarization charge of water molecules and excess ionic charge in the electrolyte. The quantity corresponding to $q$ in our model system is therefore  $q = Q_{\text{net}}/A$ with $Q_{\textrm{net}}=Q(z_e)$ defined in section \ref{sec:edlch}.  $\rho(z)$ in Eq.~\ref{Q_z_md} and the Maxwell  electric field $E_z(z)$  along the $z$ axis are related by the  Maxwell's equation
 \begin{equation}
\label{dedz}
 \frac{d E_z(z)}{dz} = 4 \pi \rho(z)
\end{equation}
 The Maxwell electric field in a neutral electrolyte ($z=z_e$)  vanishes, $E_z(z_e) = 0$.  Substituting  Eq.~\ref{dedz} in Eq.~\ref{Q_z_md} gives
 \begin{equation}
 \label{qEd}
 4 \pi Q_{\textrm{net}} = A\int^{z_e}_{-L/2} dz \frac{d E_z(z)}{dz} = -AE_\textrm{d}
\end{equation}
where $E_\textrm{d} = E_z(-L/2)$  is the field in the middle of the dielectric (insulator) layer.   $E_\textrm{d}$ in our simple system is constant (see Fig.~\ref{ics})  and finite, $E_\textrm{d} > 0$, and therefore $Q_{\textrm{net}} < 0$ in accordance with the MD results of Fig.~\ref{Qz}. The EDL is charge balanced only in the limit of a vanishing $E_\textrm{d}$. 

The electric field in the insulator plays an equally important role in determining the $z$  component of the $x,y$ averaged polarization density  $P_z$.  This quantity is obtained as the first moment of the charge density $\rho(z)$ introduced in  Eq.~\ref{Q_z_md}    
\begin{equation}
\label{defpz}
P^{\textrm{cell}} = \frac{1}{L}\int_{-L/2}^{+L/2} \rho(z) zdz 
\end{equation}
We have added the superscript ``cell'' to distinguish the polarization of Eq.~\ref{defpz} from an extended definition of polarization introduced later (Also the Cartesian component index $z$ has been suppressed).  Substituting  Eq.~\ref{dedz} followed by partial integration yields 
\begin{equation}
P^{\textrm{cell}} = \frac{zE_z(z)}{4\pi L }\Big\vert^{+L/2}_{-L/2} 
 - \frac{\bar{E}}{4\pi }
\end{equation}
where $\bar{E}$ is the average of $E_z$ over the cell
\begin{equation}
\label{barE}
\bar{E}=\frac{1}{L}\int^{+L/2}_{-L/2} E_z(z')dz' 
\end{equation}

Because of the periodicity  of the system geometries Figs.~\ref{ecs} and \ref{ics} we can set $E_z(-L/2)=E_z(+L/2)=E(L/2)$   yielding
\begin{equation}
\label{cellpolE}
 4 \pi P^{\textrm{cell}} = E(L/2)- \bar{E}
\end{equation}
Similarly $\bar{E}$, the $k=0$ component of the Maxwell field, vanishes under PBC leaving us with 
\begin{equation}
\label{cellpoltot}
 4 \pi P^{\textrm{cell}} = E(L/2)
\end{equation}

Eqs.~\ref{cellpoltot} is the supercell manifestation of the infamous ambiguity in the definition of polarization for extended periodic solids.  Polarization depends on where we draw the supercell boundary. For the system in our study, the choice is between ECS (Fig.~\ref{ecs}) and ICS boundaries (Fig.~\ref{ics}). For an ECS, $E(L/2)=E_{\textrm{d}}>0$. An ECS of finite length $L$ has a finite polarization. In contrast, for an ICS, $E(L/2)=E_{\textrm{e}}$ where $E_{\textrm{e}}$ is the field in the electrolyte. $E_{\textrm{e}} = 0 $ and hence the cell polarization of an ICS is always zero. This is a counter intuitive and most likely nonphysical result. In the next sections we will see how the consistency between the ECS and ICS representation can be restored, first for a continuum model (section \ref{sec:stern}) and then for the atomistic system of Figs~\ref{ecs} and \ref{ics}.   

Both the net EDL charge and dipole are effectively determined by $E_\textrm{d}$ which in our periodic model is finite contrary to the field in a semi-infinite dielectric (see also Ref.~\citenum{Noguera:2008rpp}). This suggests that if  $E_\textrm{d}$ can be cancelled out by some modification of the system, we would have constructed a supercell model for an interface between an electrolyte and a macroscopic dielectric.  In the following, we will show that this can be achieved by applying a finite electric field. Under these conditions  $E(L/2)=0$ both in the ECS and ICS view but now $\bar{E}$ of Eq.~\ref{barE} is finite.  This may appear at first to violate periodic boundary conditions because this would imply a net potential difference between supercell boundaries.  However, the SSV scheme is compatible with supercell geometries as shown by Vanderbilt and coworkers\cite{Stengel:2009cd, Stengel:2009prb}.  This will be demonstrated first for a Stern-like model which will be subsequently generalized to the atomistic model. This generalization is not straight forward  and will require a non-trivial adjustment of the definition Eq.~\ref{defpz} of polarization.

\section{Continuum model} \label{sec:stern}

\subsection{Stern model of parallel EDLs} \label{sec:polstern}
The  traditional approach to modelling of EDL's in colloid science is based on macroscopic Maxwell theory~\cite{Westall:1980cols,Israelachvili:2011Is}. This approach is similar to the continuum modelling of EDL's in electrochemistry~\cite{Schmickler:2010Sc,Stern:1924St} but the  focus is on fields rather than potentials.  A basic Stern model, as used in colloid science, describes the EDL as a set of charged planes separated by uniform dielectric media each with their own dielectric constant\cite{Westall:1980cols}.  The model applies to the limit of high ionic strength in which the (inverse) capacitance of diffuse layers in the electrolyte can be ignored.

\begin{figure} [h]
\includegraphics[width=1.00\columnwidth]{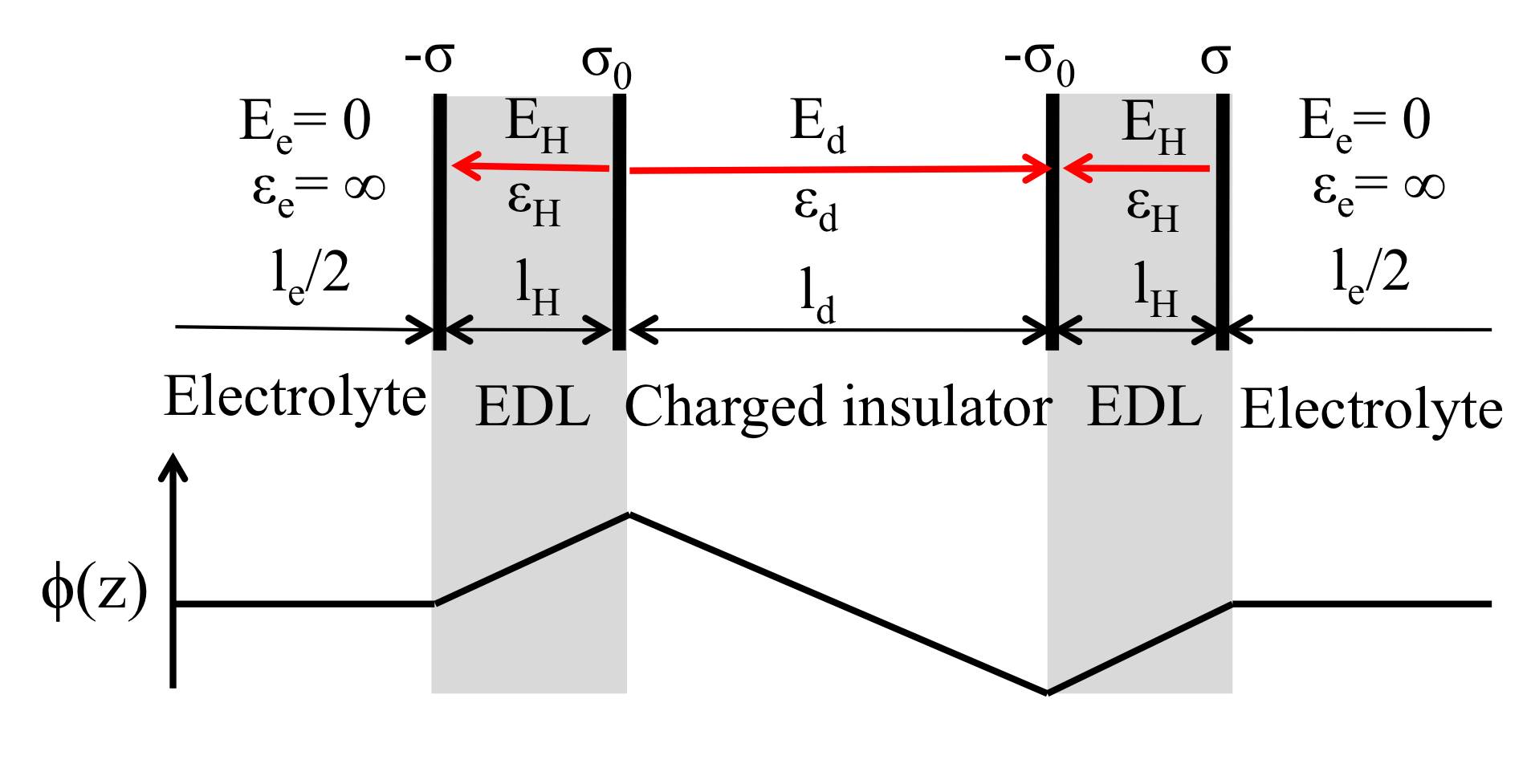}% Here is how to import EPS art
\caption{\label{sternics} Diagram of the Stern model for the atomistic ICS system of  Fig.~\ref{ics}.  $\sigma_0$ is the fixed surface charge density of the rigid walls.  $\sigma$ represents the variable surface  density of counter charge supplied by the electrolyte. The red arrows indicate the direction of the electric fields assuming $\sigma_0>0$.   $\phi(z)$ is the corresponding electrostatic potential profile.}
\end{figure}

 The minimal Stern model for our system is shown  in Fig.~\ref{sternics}, which is the continuum counterpart of Fig.~\ref{ics}. It is a quadruple layer Stern model. For simplicity, the two EDLs are considered to be equivalent. In each EDL, the dielectric constant is $\epsilon_\text{H}$, the length is $l_\text{H}$, the field inside is $E_\text{H}$.  The fixed surface charge density $\sigma_0$ is compensated by a charge density $-\sigma$ on the outer Helmholtz plane. $\sigma$ is not necessarily the same as $\sigma_0$. $\sigma$ a variable to be determined by solving the Maxwell equations. What is left of the electrolyte  is assumed to be a perfect ionic conductor of length $l_e$ and dielectric constant $\epsilon_\text{e} = \infty$. This region is therefore electric field free, $E_\text{e}=0$. The dielectric constant of the insulator is $\epsilon_\text{d}$, its length is $\l_\text{d}$ and the field inside is $E_\text{d}$. The layer widths add up to the supercell length 
\begin{equation}
\label{L2EDL}  
L=2l_\text{H}+l_\text{d}+l_\text{e}
\end{equation}

\begin{figure} [h]
\includegraphics[width=1.00\columnwidth]{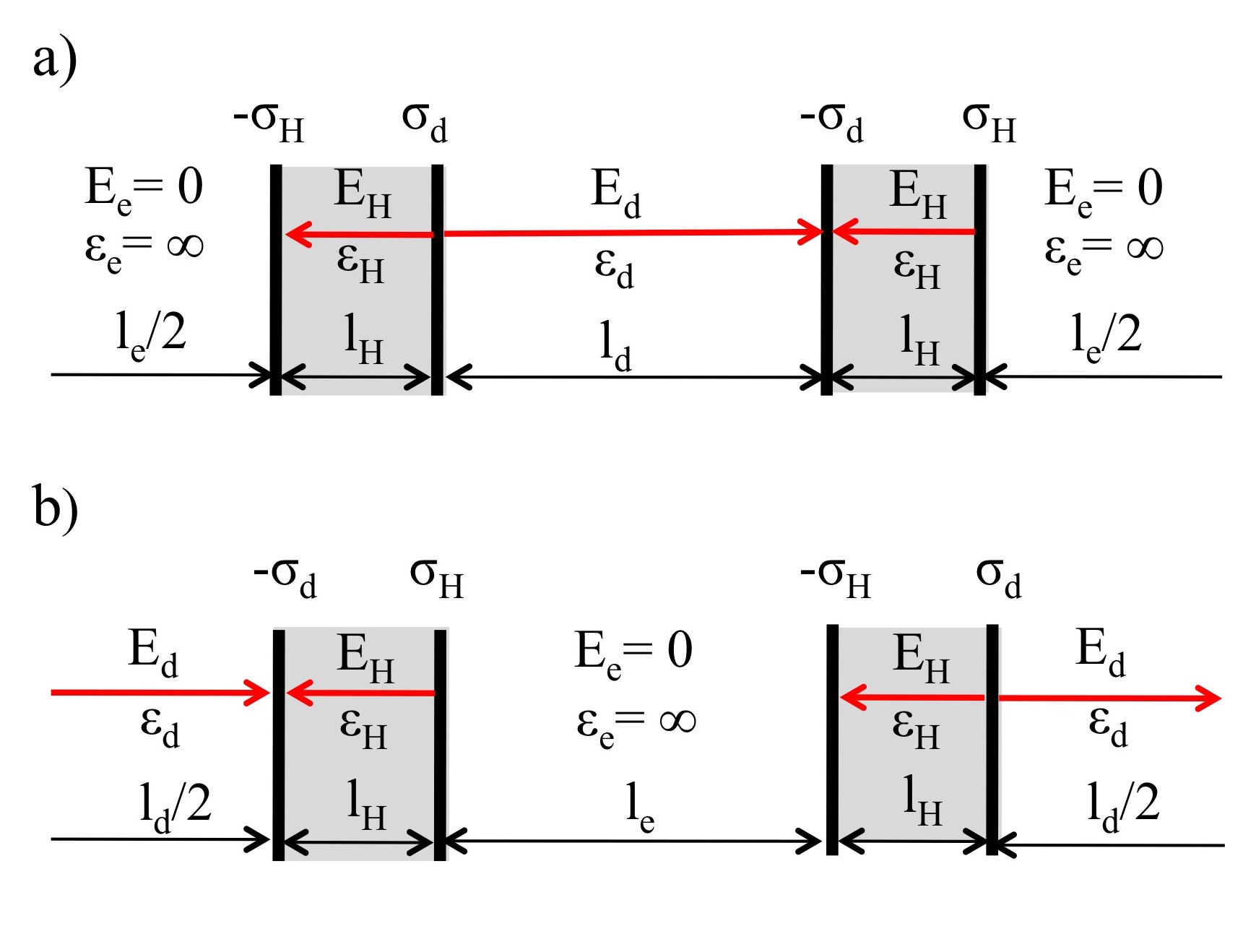}% Here is how to import EPS art
\caption{\label{sigmaE} The total charge  $\sigma_{\textrm{H}}$ and $\sigma_{\textrm{d}}$ at the Helmholtz layer-electrolyte respectively Helmholtz layer-insulator interface: a) The insulator centred supercell (ICS), and b) The electrolyte centred supercell (ECS). $\sigma_{\textrm{H}} = E_{\textrm{H}}/4\pi$ and  $\sigma_{\textrm{d}} =(E_{\textrm{d}}+E_{\textrm{H}})/4\pi$ (see Maxwell interface relations Eq.~\ref{maxwellH} and Eq.~\ref{maxwelld}).}
\end{figure}

Eq.~\ref{qEd} for the net EDL charge should directly apply with $\rho(z)$ replaced by delta functions representing surface charge densities.  However it is instructive to re-derive this equation from the interface formulation of Maxwell's law Eq.~\ref{dedz} relating the total surface charge density of a plane to the discontinuity in the Maxwell electric field. The net charge on the insulator plane (see Fig.~\ref{sigmaE}) is
\begin{equation}
\label{sigmad}
-\sigma_{\textrm{d}} = -\sigma_0 + P_{\textrm{H}} +P_{\textrm{d}} 
\end{equation}
where $P_{\textrm{H}}$ is the polarization in the Helmholtz layer and $P_{\textrm{d}}$ is the polarization in the insulator slab. $\sigma_d$ is related  to the electric fields on either side of the fixed charge plane as 
\begin{equation}
\label{maxwelld}
 4 \pi \sigma_{\textrm{d}} = E_{\textrm{d}} + E_{\textrm{H}}
\end{equation}
with the field $ E_{\textrm{H}}$ and $E_{\textrm{d}}$ pointing along the red arrows in Fig.~\ref{sternics}. Note that  $ E_{\textrm{H}}$ and  $E_{\textrm{d}}$ are defined as pointing in opposite direction. $\sigma_{\textrm{d}}$ of Eq.~\ref{maxwelld} is indicated in Fig.~\ref{sigmaE}.
The total counter charge on the plane separating the Helmholtz layer from the electrolyte is
\begin{equation}
\label{sigmaH}
\sigma_{\textrm{H}} = \sigma - P_{\textrm{H}}
\end{equation}
which is directly proportional to $E_{\textrm{H}}$ because the field $E_{e}$ in the electrolyte is zero.
\begin{equation}
\label{maxwellH}
4 \pi \sigma_{\textrm{H}} =  E_{\textrm{H}}
\end{equation}
The net EDL charge density $q$ is the sum of $\sigma_{\textrm{d}}$ and $\sigma_{\textrm{H}}$. 
Adding Eqs.~\ref{sigmad} and \ref{sigmaH} we find
\begin{equation}
  Q_{\textrm{net}} =  A(-\sigma_{\textrm{d}}  + \sigma_{\textrm{H}}) = A(-\sigma_0 + \sigma + P_{\textrm{d}})
\end{equation}
which according to  Eqs.~\ref{maxwelld} and \ref{maxwellH} is equal to
\begin{equation}
\label{qEDL}
 Q_{\textrm{net}}  = -AE_{\textrm{d}}/4 \pi
\end{equation}
reproducing Eq.~\ref{qEd}.

To obtain an expression for the polarization of the Stern model, it would seem that we simply have to evaluate the moment of the charge sheets in Fig.~\ref{sigmaE} distinguishing between the ICS (Fig.~\ref{sigmaE}a) and ECS (Fig.~\ref{sigmaE}b) geometry. However, formally, the cell polarization $ P^{\textrm{cell}}$ as defined in Eq.~\ref{defpz} in terms of a volume dipole moment is only part of the supercell polarization. Application of the divergence theorem yields an additional term due to the polarization surface charge density at the boundary of the cell\cite{Landau:1960v8,Martin:1974prb}. The volume integral of the  $z$ component of the polarization should be written as
\begin{equation}
\label{gausspol}
 \int_{\mathrm{cell}} d\mathbf{r} P_z(\mathbf{r}) =  \int_{\mathrm{cell}} d\mathbf{r} z \rho(\mathbf{r}) + \int_{\mathrm{surface}} ds z \mathbf{n} \cdot \mathbf{P}(\mathbf{r}) 
\end{equation}
where $\rho(\mathbf{r}) = - \nabla \cdot \mathbf{P}(\mathbf{r})$  is the volume polarization charge density as before. The second term is minus the surface polarization charge density of the supercell ($\mathbf{n}$ is the outward normal to cell boundary surface).  

For microscopic point charge distributions (as in the SPC model shown in Fig.~\ref{ecs} and Fig.~\ref{ics}) the surface term is usually omitted. It should always be possible to construct a boundary surface avoiding all point charges.  This is why the surface term was ignored in section \ref{sec:edielec}. For electronic charge densities in an extended system (in particular semiconductors) the surface term is finite\cite{Martin:1974prb}. This observation by Richard Martin was one of the key arguments convincing the solid state community that a new approach to polarization was required which eventually led to the modern theory of [polarization.  Similarly the surface term must be taken into account  for the continuum charge density of the periodic Stern model. For our effectively one-dimensional system geometry Eq.~\ref{gausspol} can be formulated as            
\begin{equation}
\label{defpzs}
P = P^{\textrm{cell}} + P(L/2)  
\end{equation}
where $P^{\textrm{cell}}$ is the volume polarization of Eq.~\ref{defpz} (The superscript for the Cartesian $z$ component has again be omitted). 

The evaluation of Eq.~\ref{defpzs} will be carried out first for the electrolyte centred supercell (ECS) of Fig.~\ref{sigmaE}b. Substituting Eqs.~\ref{maxwelld} and \ref{maxwellH} for the plane charges and find 
\begin{eqnarray}
 P^{\textrm{cell}}_{\textrm{ECS}}   &= & \frac{1}{L} \left( \frac{(E_{\textrm{H}} + E_{\textrm{d}})}{4 \pi } (l_{\textrm{e}} + 2l_{\textrm{H}})
 -\frac{E_{\textrm{H}}}{4 \pi } l_{\textrm{e}} \right) \nonumber \\
   & = & \frac{1}{4 \pi L} \left(  E_{\textrm{d}} (l_{\textrm{e}} + 2l_{\textrm{H}}) 
     + E_{\textrm{H}}  2  l_{\textrm{H}} \right) \nonumber \\
  & = & \frac{1}{4 \pi L} \left( E_{\textrm{d}}  L 
  + ( E_{\textrm{H}}  2  l_{\textrm{H}} -  E_{\textrm{d}} l_{\textrm{d}}) \right)  \label{pECS}
\end{eqnarray}
where we have used Eq.~\ref{L2EDL}. 
The right hand side (rhs) of Eq.~\ref{pECS} can be simplified by rewriting in terms of $\bar{E}$. The  electric field is piece wise uniform and the field integral of Eq.~\ref{barE} becomes  a sum of potential changes over the various dielectric subsystems 
\begin{equation}
\label{field_pbc1}
-E_\text{H}2l_\text{H} + E_\text{d}l_\text{d} + E_\text{e}l_\text{e} = \bar{E}L
\end{equation}
Eq.~\ref{field_pbc1} can be regarded as an application of Volta's law.  Since $E_\text{e}=0$, the above equation is reduced to
\begin{equation}
\label{field_pbc2}
 E_\text{H}2l_\text{H} - E_\text{d}l_\text{d} = -\bar{E}L
\end{equation}
Replacing the last two terms of Eq.~\ref{pECS} then gives 
\begin{equation}
\label{pcellecs}
4\pi P^{\textrm{cell}}_{\textrm{ECS}} =E_{\textrm{d}} - \bar{E}
\end{equation}
which is the volume term in Eq.~\ref{defpzs}. 

The surface term can be obtained from a constitutive relation. The ECS is terminated in the dielectric insulator. The polarization $P_{\textrm{ECS}}(L/2)$ at the ECS boundaries is therefore proportional to the electric field $E_d$ in the dielectric according to $P_{\textrm{ECS}}(L/2) = P_d = (\epsilon_d -1 ) E_d/4 \pi$. Inserting in Eq.~\ref{defpzs} together with Eq.~\ref{pcellecs} we find
\begin{equation}
\label{psterne}
4\pi P_{\textrm{ECS}} =\epsilon_d E_{\textrm{d}} - \bar{E}
\end{equation}
Comparing to Eq.~\ref{cellpolE} we see that including the surface term replaces the macroscopic electric field $E(L/2) = E_d$ by the corresponding displacement field $D_d = \epsilon_d E_d$ enhancing the polarization.    

How will the surface term change the polarization in the ICS representation? Evaluation of the volume (cell) polarization of the ICS is simple enough. Computing the dipole moment of the charge sheets in the cell we find  
 \begin{equation}
 P^{\textrm{cell}}_{\textrm{ICS}} =  \frac{1}{L} \left( \sigma_{\textrm{d}}  l_{\textrm{d}} +
 \sigma_{\textrm{H}}(l_{\textrm{d}} + 2l_{\textrm{H}})\right) 
 \end{equation}
Substitution of Eqs.~\ref{maxwelld} and \ref{maxwellH} yields
\begin{equation}
 P^{\textrm{cell}}_{\textrm{ICS}}   =  \frac{1}{4 \pi L}
      \left( E_{\textrm{H}}2  l_{\textrm{H}}  - E_{\textrm{d}} l_{\textrm{d}} \right)  
\label{pICS}
\end{equation}
Finally inserting Eq.~\ref{field_pbc2} gives
\begin{equation}
   \label{pcellics}
4 \pi P^{\textrm{cell}}_{\textrm{ICS}} = - \bar{E}
\end{equation}
reproducing the result of section \ref{sec:edielec}. 

Without surface term there never can be any polarization in the ICS geometry under standard Ewald summation ($\bar{E}=0$). The surface term $P(L/2)$ in Eq.~\ref{defpzs} is evidently crucial.  However, the cell boundary is now a plane in the electrolyte. What is the polarization in a conductor? There is no macroscopic electric field (perfect screening). It is therefore sometimes argued that the polarization is equally zero and therefore  also the electric displacement. Here, however, we view the electrolyte as a body with infinite dielectric constant (Fig.~\ref{sternics}). $\epsilon_e = \infty$ implies not zero but  unit polarizability  and therefore $4 \pi P_e = (1 - 1/\epsilon_e ) D_e = D_e$. Accordingly, setting  $P(L/2)$ in Eq.~\ref{defpzs}  to $D_e$ we arrive at 
\begin{equation}
   \label{psternd}
4 \pi P_{\textrm{ICS}} = D_e - \bar{E}
\end{equation}
Eq.~\ref{psternd}  is of the same form as  Eq.~\ref{psterne} for the polarization of the ECS. In fact if $D_e=D_d = \epsilon_d E_d$ the conclusion would be that $P_{\textrm{ICS}} = P_{\textrm{ECS}}$. The polarization would be invariant for a change of supercell contrary to what one is led to expect from the modern theory of polarization. A change of supercell can add or subtract a so called ``polarization quantum'' to the polarization (for a more detailed discussion see section \ref{sec:dpzc}). We will show now that  $P_{\textrm{ICS}}$ and $P_{\textrm{ECS}}$ differ by $\sigma_0$ which plays the role of polarization quantum in the continuum model. 

Again in virtue of the constitutive relations we can replace  $P_{\textrm{H}}$ by  $\left(\epsilon_{\textrm{H}} - 1 \right) E_{\textrm{H}}/4 \pi$ and similarly $P_{\textrm{d}}$ by  $\left(\epsilon_{\textrm{d}} - 1 \right) E_{\textrm{d}}/4 \pi$ in Eq.~\ref{sigmad}. This gives an expression for $\sigma_\textrm{d}$ in terms of $\sigma_0$ and $E_{\textrm{H}}$. Substituting for $\sigma_\textrm{d}$  in Eq.~\ref{maxwelld} using the sign convention of Fig~\ref{sternics} we find
\begin{equation}
\label{maxwells0}
\epsilon_{\textrm{H}} E_{\textrm{H}} +\epsilon_{\textrm{d}} E_{\textrm{d}}  = 4\pi\sigma_0
\end{equation}
Then as usual defining the \emph{local} electric displacement as $D_{\textrm{H}} = \epsilon_{\textrm{H}} E_{\textrm{H}}$ and $D_{d} = \epsilon_{d} E_{d}$ we can write
\begin{equation}
\label{deltaDHd}
 D_{\textrm{H}} + D_d = 4 \pi \sigma_0
\end{equation} 
The electric displacement is discontinuous when crossing the plane of fixed charge density $\sigma_0$ separating the Helmholtz layer and insulator slab.
    
 On the other hand, continuing from the Helmholtz layer to the electrolyte $D$ remains constant. The reason is that, once we have allowed for a finite polarization in the electrolyte, the response charge $\sigma$ must be considered as the corresponding polarization charge and therefore $P_e = \sigma$ (Note we have aligned the positive direction of the electrolyte polarization with the polarization of the dielectric).  Furthermore applying the constitutive relations  to Eqs.~\ref{sigmaH} and \ref{maxwellH} gives
\begin{equation}
\label{maxwellsH}
D_{\textrm{H}}=\epsilon_{\textrm{H}} E_{\textrm{H}} = 4\pi\sigma
\end{equation}
Unlike Eq.~\ref{maxwells0} there is only a single field term because on the electrolyte side of the interface the field is zero. Then combining  with $4 \pi \sigma = 4 \pi P_e = D_e$ we obtain 
\begin{equation}
\label{deltaDHe}
D_{\textrm{H}} + D_e = 0    
\end{equation}
Finally subtracting Eq.~\ref{psternd} and \ref{psterne} using Eqs~\ref{deltaDHd} and \ref{deltaDHe} we find
\begin{equation}
\label{pics2ecs}
 P_{\textrm{ICS}} =  P_{\textrm{ECS}} + \sigma_0
\end{equation} 
A shift of the supercell by half a cell length changes the polarization by the fixed surface charge density responsible for inducing the ``built-in'' polarization, i.e.~the finite polarization at zero macroscopic field $\bar{E}$. 

The implication of Eq.~\ref{pics2ecs} is that the polarization is not unique but depends on the choice of supercell. Technically this ambiguity is the result of including the fixed plane charge $\sigma_0$ in the definition of polarization. However, more generally, Eq.~\ref{pics2ecs} can be regarded as a manifestation of a key concept of the modern theory of polarization referred to as the multivaluedness of polarization in periodic extended systems. $\sigma_0$ in Eq.~\ref{pics2ecs} can be interpreted as a ``polarization quantum'' usually written as $e/A$~\cite{King-Smith:1993prb,Resta:1994rmp,Resta:2007ch}.  Multivaluedness is even more of an issue for the atomistic model where, in addition to the fixed surface charge, the explicit mobile ions can also contribute. Further discussion is therefore deferred to section \ref{sec:gauge} and \ref{sec:dpzc} on itinerant polarization.    

\subsection{ Constant electric field $\bar{E}$ and point of ZNC} \label{sec:sternsc}
We now return to the problem of the non-zero net charge of the double layer (Eq.~\ref{qEDL}) and exploit the simplicity of the Stern model and derive an analytic expression for $E_d$. The quantities $\sigma, E_{\textrm{H}}$ and $E_{\textrm{d}}$  are treated as variables uniquely determined by solving Eqs.~\ref{field_pbc2}, \ref{maxwells0} and \ref{maxwellsH} with the  insulator surface charge $\sigma_0$ and average field  $\bar{E}$ as control parameters. This leads to
 \begin{eqnarray}
\label{EHc}
\frac{\epsilon_{\textrm{H}}E_{\textrm{H}}}{4 \pi} &=& \sigma =
\frac{ \sigma_0/ C_{\textrm{d}} - \bar{E} L}{ 2/C_{\textrm{H}} + 1/C_{\textrm{d}}} \\
\label{Edc}
\frac{\epsilon_{\textrm{d}}E_{\textrm{d}}}{4 \pi} &=& \sigma_0 -\sigma =
\frac{ 2 \sigma_0/ C_{\textrm{H}} + \bar{E} L}{ 2/C_{\textrm{H}} + 1/C_{\textrm{d}}}
\end{eqnarray}
where $C_{\textrm{H}}$ is the capacitance of the Helmholtz layer and $C_{\textrm{d}}$ the capacitance of the dielectric material given by 
\begin{equation}
\label{CHCd}
  C_{\textrm{H}} = \frac{\epsilon_{\textrm{H}}}{4 \pi l_{\textrm{H}}}, \qquad
 C_{\textrm{d}} = \frac{\epsilon_{\textrm{d}}}{4 \pi l_{\textrm{d}}}
\end{equation}

The  product $-\bar{E}L$ in Eqs.~\ref{EHc} and \ref{Edc}  is interpreted as the potential difference across the length of the supercell.  Treated as  a control parameter $V_{\textrm{ext}}= -\bar{E}L$ acts as an external bias. Accordingly, the net EDL  surface charge (Eq.~\ref{qEDL}) can be expressed as:
\begin{equation}
\label{Qnet_stern_E}
Q_{\textrm{net}}(V_{\textrm{ext}}) =\frac{A}{\epsilon_\textrm{d}} \left (\frac{  V_{\textrm{ext}} - 2 \sigma_0/ C_{\textrm{H}}}{ 2/C_{\textrm{H}} + 1/C_{\textrm{d}}} \right)
\end{equation}
with the opposite charge for the opposite EDL. Under PBC conditions $V_{\textrm{ext}}=0$ (short circuit). The net charge plotted in Fig.~\ref{Qz}b is predicted to scale as $1/l_{\textrm{d}}$  for increasingly large vacuum layer width $l_{\textrm{d}}$:
\begin{equation}
\label{Qnet_stern}
Q_{\text{net}} =-\frac{\sigma_02l_\text{H}A}{l_\text{d}\epsilon_\text{H}
  + 2l_\text{H}\epsilon_\text{d}}
\end{equation}
 The simulation results are consistent with Eq.~\ref{Qnet_stern}. The Helmholtz EDL will approach charge balance when $C_{\textrm{d}} \rightarrow 0$ in the limit $l_\text{d}\to\infty$ .  

The way out is also suggested by Eq.~\ref{Qnet_stern_E}. The argument can be turned around asking the question what would be the value of $\bar{E}$ for charge compensated EDL's, or in the terminology of the introduction, the state of zero net charge (ZNC). Setting $\sigma = \sigma_0$ in Eq.~\ref{Qnet_stern_E} we find
\begin{equation}
\label{pzc} 
V_{\textrm{znc}} = -L \bar{E}_{\textrm{znc}} = \frac{2 \sigma_0}{C_{\textrm{H}}}  
\end{equation}
Full screening of the fixed charge $\sigma_0$ requires assistance of an  external field, or equivalently a bias potential.   The smaller the capacitance $C_{\textrm{H}}$ of the Helmholtz layer the larger field is needed. The Helmholtz layer impedes the screening power of the conductor by inserting a dielectric layer between the conductor and the fixed charge similar to the dead layer in ferroelectric nanocapacitors~\cite{Stengel:2006kv,Stengel:2009go}.

The potential at the  point of ZNC is for given surface charge $\sigma_0$  fully determined by the capacitance of the Helmholtz layer. It is independent of the dielectric constant of the insulator slab. Most importantly $V_{\textrm{znc}}$  is not an extensive property scaling linearly with system size dimensions such as the width  $l_{\textrm{d}}$ or cell length $L$ (of course there might be a weaker dependence due to finite size errors).  We will take these observations as a justification for the claim that under ZNC conditions the EDL in our (still small) periodic model can be compared to the EDL in a macroscopic system for the same surface charge density.

A  Stern model  is a linear continuum theory and one expects a linear dependence of polarization on the fixed charge $\sigma_0$ and macroscopic electric field $\bar{E}$.   Investigating again  the ECS first  the equation of state (EOS) for $P_{\textrm{ECS}}$ is written in the general form 
\begin{equation}
\label{eosE}
 4 \pi P_{\textrm{ECS}} \left(\sigma_0, \bar{E} \right) = \gamma_E \sigma_0 +  \left(\bar{\epsilon} - 1 \right)\bar{E}
\end{equation}
$\bar{\epsilon}$ plays the role  of the global dielectric permittivity of the heterogeneous system. $\gamma$ is a generalized inverse capacitance.   Substituting Eq.~\ref{Edc} into Eq.~\ref{psterne} we obtain 
\begin{eqnarray}
\label{bareps}
\bar{\epsilon}  & = & \frac{ 4 \pi  L}{2/C_{\textrm{H}}+1/C_{\textrm{d}}}
  \\[6pt]
\label{gammae}
 \gamma_E & = &  \frac{ 8 \pi/C_{\textrm{H}}}{2/C_{\textrm{H}}+1/C_{\textrm{d}}}
 = \frac{2 \bar{\epsilon}}{LC_{\textrm{H}}}
 \end{eqnarray}
 Inserting Eq~\ref{pzc} in Eq.~\ref{eosE}  the EOS at the point of ZNC reduces to
\begin{equation}
\label{polpzc}
  4 \pi P_{\textrm{znc}} = - \bar{E}_{\textrm{znc}}
\end{equation} 
Eq.~\ref{polpzc} states that the point of ZNC corresponds to an open circuit capacitor without any depolarization by the metal electrodes. 

As we saw in section \ref{sec:polstern} the global polarization of continuum model differs by $\sigma_0$  when evaluated for the electrolyte or insulator supercell (Eq.~\ref{pics2ecs}). Thus, the EOS of the ICS is obtained by shifting EOS of Eq.~\ref{eosE} by $\sigma_0$.
\begin{equation}
\label{ICSeosE}
 4 \pi P_{\textrm{ICS}} \left(\sigma_0, \bar{E} \right) = \gamma_E  \sigma_0 +  \left(\bar{\epsilon} - 1 \right)\bar{E} + 4\pi\sigma_0
\end{equation}
Subtracting the polarization quantum $\sigma_0$ from $P_{\textrm{ICS}}$, Eq.~\ref{polpzc} is recovered for the ICS as well.

\subsection{ Constant electric displacement $\bar{D}$ and capacitance} \label{sec:sternsc_D}

Stengel and coworkers in their pioneering work on constant D methods  point out that using the electric displacement field $\mathbf{D}$ rather than the macroscopic electric field $\mathbf{E}$ as control field can have certain advantages\cite{Stengel:2009cd,Stengel:2009prb,Stengel:2009natm,Stengel:2009prr,Stengel:2011prl,Cazorla:2012prb}. Implemented for our quasi one dimensional slab geometry we introduce a global displacement field $\bar{D}$ defining it as the thermodynamic conjugate to the average  macroscopic electric field $\bar{E}$ of Eq.~\ref{barE}. $\bar{D}$ is related to $\bar{E}$ as 
\begin{equation}
\label{defdbar}
\bar{D}= \bar{E} + 4 \pi P
\end{equation} 
For our periodic continuum model  $P$ is the polarization (volume plus surface term) defined in Eq.~\ref{defpzs}. We have again suppressed the Cartesian coordinate $z$. 

Eq.~\ref{defdbar} suggests that $\bar{D}$ is the displacement field generated by external charge on the virtual electrodes representing the electric boundary conditions at infinity. Treated as a variable open circuit constraint $\bar{D}$ can be used to probe the electric equation of state which turned out to be very instructive for ferro-electric systems\cite{Stengel:2009cd,Stengel:2009prb}. Note, however, that in our system $\bar{D}$ is only a component of the displacement field because, as shown by Eq.~\ref{deltaDHd}, the displacement field is not uniform (recall the change in the reference of the electric field in the Helmholtz layer as shown in Fig.~\ref{sternics}). 

In section  \ref{sec:sternsc} we derived an equation  $P(\sigma_0, \bar{E})$ for the polarization as function of insulator surface charge $\sigma_0$  and average electric field $\bar{E}$.  This equation can be transformed to $P(\sigma_0, \bar{D})$ form by switching from  $\bar{E}$ to $\bar{D}$ using Eq.~\ref{defdbar}. Substituting in Eq.~\ref{eosE} the resulting EOS for the electrolyte centred supercell  can be written as
\begin{equation}
\label{eosD}
4\pi P_{\textrm{ECS}} \left(\sigma_0, \bar{D} \right)  =\gamma_D\sigma_0  + 
\left(1 -\frac{1}{\bar{\epsilon}} \right)\bar{D}
\end{equation}
$\bar{\epsilon}$ is the effective permittivity as given by Eq.~\ref{bareps}.   $\gamma_D$ is given by
\begin{equation}
\label{gammad}
  \gamma_D = \frac{\gamma_E}{\bar{\epsilon}} = \frac{2}{LC_{\textrm{H}}}
\end{equation}
The coefficient of $\bar{D}$ in Eq.~\ref{eosD}  can be regarded as  the supercell polarizability.  

Comparing Eq.~\ref{gammad} to Eq.~\ref{gammae} we see that all dependence on the dielectric constant of the insulator slab has been be eliminated.  This is a major gain. It means that the  response of polarization to a change in the fixed charge density at constant electric displacement is a direct probe of the capacitance of the Helmholtz layer. Formulated in a more thermodynamic form we can write 
\begin{equation}
\label{deltaP_D}
\left( \frac{ \partial L P_{\textrm{ECS}}}{ \partial \sigma_0}\right)_{\bar{D}} = \frac{2}{4\pi C_{\textrm{H}}} 
\end{equation}
Even more important, we are  now able to estimate the Helmholtz capacitance $C_{\textrm{H}}$ without having to locate the point of zero charge. To make this statement explicit we substitute Eq.~\ref{deltaP_D} in Eq.~\ref{pzc} giving
\begin{equation}
\label{pzc2}
V_{\textrm{znc}} = 4\pi \sigma_0 
\left( \frac{ \partial L P_{\textrm{ECS}}}{ \partial \sigma_0}\right)_{\bar{D}} 
\end{equation}
Note that $LP$  is the dipole moment of the cell including the surface charge at the boundaries. At ZNC the dipole moment is entirely determined by the interface double layers and should therefore be independent of $L$. If we are going to use Eqs.~\ref{deltaP_D} to compute the capacitance for atomistic systems, we must assume, of course, that the linear response approximation remains valid.

What would be the corresponding dependence of the polarization on the displacement field in the ICS geometry? Formally this EOS should follow from substituting Eq.~\ref{defdbar} in Eq.~\ref{ICSeosE}. However, the extra $\sigma_0$ term in Eq.~\ref{ICSeosE} contributes to the derivative Eq.~\ref{deltaP_D} which would therefore no longer be proportional to the inverse capacitance. Moreover, consistency requires that the polarization defining $\bar{D}$ for the ICS is $P_{\textrm{ICS}}$ of Eq.~\ref{pics2ecs} which would add a further $\sigma_0$ dependence.  At the end of section \ref{sec:polstern} we argued that the $\sigma_0$ term in Eq.~\ref{pics2ecs} must be understood in the context of the modern theory of polarization. It is a consequence of a change of representation (the technical term is ``branch'') of the multivalued polarization and should ultimately have no effect on physical observables such as capacitance.     

To reconcile the electric EOS for the ICS and the EOS of Eq.~\ref{eosD} for the ECS we first note that the displacement field is subject to the same multivaluedness as the polarization. The macroscopic field $\bar{E}$ (effectively the potential difference over a cell length) is unique and should therefore be invariant for a change of supercell. This is achieved by defining a separate displacement field specific to the ICS according to 
\begin{equation}
\label{dics2ecs}
 D_{\textrm{ICS}} =  D_{\textrm{ECS}} + 4\pi\sigma_0
\end{equation}
With the shifted polarization Eq.~\ref{pics2ecs} and shifted displacement field Eqs.~\ref{dics2ecs}, Eq.~\ref{ICSeosE} yields
\begin{equation}
\label{ICSeosD}
4\pi P_{\textrm{ICS}} \left(\sigma_0, \bar{D}_{\textrm{ICS}} \right)  =\gamma_D\sigma_0  + \left(1 -\frac{1}{\bar{\epsilon}} \right)\bar{D}_{\textrm{ICS}} + 4\pi\sigma_0
\end{equation}
The conclusion is that the EOS for polarization  in the ICS and ECS can be made to coincide by subtracting the polarization quantum $\sigma_0$ from both the polarization and displacement field. This simple transformation can be carried out because $\sigma_0$ is a known parameter. As a result, Eq.~\ref{deltaP_D} and Eq.~\ref{pzc2} can be used for the ECS as well as the ICS. In section \ref{sec:deltaP} we will see how the same applies to the atomistic system with the difference that we may now need the add or subtract a multiple of the elementary polarization quantum $e/A$.

\section{Finite field methods} \label{sec:fields}

\subsection{Electric hamiltonians and thermodynamics} \label{sec:ssv}
The Stern continuum model of section \ref{sec:stern} provides insight in the origin of the phenomenon of uncompensated EDL's under PBC (Fig.~\ref{Qz}). It also suggests two methods to fix the problem, namely  scanning the macroscopic electric field $\bar{E}$ to locate the point of zero net charge or imposing $\bar{D}=0$ (open circuit) electrical boundaries.  Our insulator-electrolyte interface model is a \emph{periodic} heterogeneous system.  $\bar{E}$ is therefore an average over a supercell. The  SSV finite field method is precisely designed for such a system\cite{Stengel:2009cd,Stengel:2009prb}. 

From a traditional molecular dynamics (MD) perspective the SSV method can be seen as an example of an extended Hamiltonian and can be readily generalized to finite temperature as we have demonstrated in a recent application to the simulation of finite temperature polarization fluctuations in polar liquids~\cite{Zhang:2015ms, Zhang:2016ho}. Finite electric field methods have been used before in classical simulation of aqueous systems\cite{Yeh:1999hb,Vega:2011prl,vanGunsteren:2011jctc}. The SSV Hamiltonian is different because of several special features which will be briefly summarized below.

 The SSV constant macroscopic field Hamiltonian $H_E$, adapted to the geometry and notation used in the present paper, is written as
\begin{equation}    
\label{fhmd}
H_E\left( v, \bar{E} \right) = H_{\mathrm{PBC}}(v)- \mathit{\Omega} \bar{E} P(v)
\end{equation}
$H_{\textrm{PBC}}(v)$ is the Hamiltonian  as defined by the force field model. $v= (\mathbf{r}^N ,\mathbf{p}^N)$ stands for the collective  momenta and position coordinates of the $N$ particles in the system. We have appended a subscript PBC to indicate that the electrostatic energies and forces are computed using standard Ewald summation (``tin foil'' boundary conditions, no surface terms). An electric field with magnitude $E_z = \bar{E}$ is applied along the $z$ axis. $P = P_z$ is the polarization. Again, the Cartesian component index $z$ has been suppressed. $\bar{E}$ is treated as a parameter, which is why it was added as a formal argument of the electric enthalpy hamiltonian $H_E\left(v, \bar{E} \right)$.  $\mathit{\Omega}$ is the volume of the MD cell. $\mathit{\Omega} = LA$ in terms of the cell length $L$ and perpendicular cross section $A$ as introduced earlier. 

The Hamiltonian of Eq.~\ref{fhmd} superficially resembles the Hamiltonians which have been applied in finite field simulation in force field MD\cite{Yeh:1999hb,Vega:2011prl}. However $\bar{E}$ is the MD cell average of the screened macroscopic electric field, not the applied electric $E_0$ as would be present in absence of the system. The coupling of polarization to $\bar{E}$ rather than $E_0$ is a consequence of using the Ewald sum for the evaluation of the long range electrostatic interactions(for further comment on this issue see our paper of Ref.~\citenum{Zhang:2015ms}). A second equally crucial point is that $P$ in Eq.~\ref{fhmd} is the itinerant polarization of the system as obtained from a time integral of the volume integral of current. For a system of polar molecules the itinerant polarization is the sum of the molecular dipoles moments which is what is normally used in numerical simulation of polar liquids. It also has been appreciated in the physical chemistry literature that using the cell polarization $P^{\textrm{cell}}$ in Eq.~\ref{defpz} would lead to erroneous results. For ionic solutions, the question of the correct definition of polarization is equally critical. Further explanation will be deferred to Section \ref{sec:gauge}. 

The extended Hamiltonian $H_E\left(v, \bar{E} \right)$ of Eq.~\ref{fhmd} generates a field dependent partition function
\begin{equation}
\label{zfmd}
Z_E = \int d\mathbf{p}^N d\mathbf{r}^N \exp \lbrack - \beta H_E\left( v, \bar{E} \right)  \rbrack
\end{equation}
 $\beta = 1/k_{\mathrm{B}}T $ is the inverse temperature. The combinatorial prefactor $1/(h^{3N} N!)$ has been omitted.  The corresponding free energy is 
\begin{equation}
\label{fmd} 
F\left( \bar{E}\right) = - k_{\mathrm{B}}T \ln Z_E
\end{equation}
The derivative of $F\left( \bar{E}\right)$ of Eq.~\ref{fmd} gives the expectation value of polarization denoted by $ \langle P(\bar{E}) \rangle  = \langle P_z(\bar{E}) \rangle$
\begin{equation}
\label{dfde}
\frac{d F}{d \bar{E}} = -\mathit{\Omega} \langle P(\bar{E})\rangle  
\end{equation}    
The polarization function $\langle P(\bar{E}) \rangle$ was referred to in section \ref{sec:sternsc} as the electric EOS. Eq.~\ref{eosE} gives its form for the continuum model.

The complementary SSV Hamiltonian for simulation under constant electric displacement $D_z$ in the $z$ direction is defined as
\begin{equation}
\label{uvdb}
H_D\left(v ,\bar{D} \right) = H_{\mathrm{PBC}}(v)+ \frac{\mathit{\Omega}}{8 \pi}
  \left(\bar{D} - 4 \pi P(v) \right)^2
\end{equation}
where we have again changed notation from $D_z$ to $\bar{D}$. The displacement $\bar{D}$ is related to the electric field $\bar{E}$ according to 
\begin{equation}
\label{ddef}
\bar{D}=\bar{E}+4\pi P
\end{equation}
This equation is the itinerant polarization counterpart of Eq.~\ref{defdbar} for the continuum model.

 The corresponding $\bar{D}$ dependent free energy is obtained from the partition function
\begin{equation}
\label{umd}
U\left( \bar{D}\right) = - k_{\mathrm{B}}T \ln Z_D
\end{equation}
with
\begin{equation}
\label{zumd}
Z_D = \int d\mathbf{p}^N d\mathbf{r}^N \exp \left[ - \beta \, H_D\left(v ,\bar{D} \right) \right]
\end{equation}
$U(\bar{D})$  and $F(\bar{E})$ are both  (Helmholtz) free energies with respect to temperature. They have a different status as electric thermodynamic potentials.  $U(\bar{D})$ is an electric internal energy, $F(\bar{E})$ an electric enthalpy.  As explained in the supplementary information of Ref.~\citenum{Stengel:2009cd} the relation between $U(\bar{D})$ and $F(\bar{E})$ is similar to that between thermodynamic conjugate potentials, but not quite. While
\begin{equation}
\label{dudd}
\bar{E}  =  \frac{4\pi}{\mathit{\Omega}} \frac{d U}{d \bar{D}}
\end{equation}
the Legendre transform of $U\left( \bar{D}\right)$ 
\begin{equation}
\tilde{F}\left(\bar{E}\right) = U\left( \bar{D}\right) - 
 \frac{\Omega}{4 \pi} \bar{E} \bar{D}
\end{equation}
differs from $F(\bar{E})$ by (minus) the field energy
\begin{equation}
\tilde{F}\left(\bar{E}\right) = F\left( \bar{E}\right) - \frac{\Omega}{8 \pi} \bar{E}^2
\end{equation}
Hence, the electric field derivative of  $\tilde{F}(\bar{E})$, not  ${F}(\bar{E})$, generates $\bar{D}$
\begin{equation}
\label{dtfde}
\bar{D}  = - \frac{4\pi}{\mathit{\Omega}} \frac{d \tilde{F}}{d \bar{E}}
 \end{equation}
The function $\tilde{F}\left(\bar{E}\right)$ is therefore the proper electric enthalpy conform to the definition in Landau and Lifshitz\cite{Landau:1960v8}.

\subsection{Itinerant polarization and ionic solutions}
\label{sec:gauge}

As pointed out in textbooks on solid state physics there is a fundamental question concerning the definition of polarization in solids of infinite extension.  The dipole moment of a unit cell is not unique but depends on how the solid is partitioned in unit cells. This ambiguity was resolved by Vanderbilt and Resta by accepting it as a fundamental property of macroscopic polarization in extended system. The macroscopic polarization according to the modern theory of polarization is a multivalued quantity and only relative values of polarization matter~\cite{King-Smith:1993prb,Resta:1994rmp,Resta:2007ch}. A change in polarization is \emph{defined} as the time integral of the volume integral of the transient current density $\mathbf{j}$:
\begin{equation}
\label{defp2}
\Delta \mathbf{P} = \mathbf{P}(t) - \mathbf{P}(0) = \frac{1}{\mathit{\Omega}}\int^{t}_{0} dt  \int_{\textrm{cell}} d\mathbf{r} \, \mathbf{j}(\mathbf{r},t)
\end{equation}
where $\mathit{\Omega}$ is again the volume of the periodic cell. 

Eq.~\ref{defp2} led to the famous Berry phase expression for electronic polarization\cite{King-Smith:1993prb,Resta:1994rmp,Resta:2007ch}.  Applied to simple point charge systems, the volume integral can be carried out immediately yielding a sum over the velocities of the particles in the MD cell: 
\begin{equation}
\label{current}
\int_{\textrm{cell}} d\mathbf{r} \, \mathbf{j}(\mathbf{r},t)
 = \sum_i^{\textrm{cell}}  q_i \dot{\mathbf{r}}_i (t)
\end{equation}
$q_i$ is the charge of particle $i$. Integrating over time we find
\begin{equation}
\label{piter}
 \mathbf{P}(t) =  \mathbf{P}(0) +  \frac{1}{\mathit{\Omega}}
\sum_i^{\textrm{cell}} q_i \Delta \mathbf{r}_i(t)
\end{equation}
$\Delta \mathbf{r}_i(t) = \mathbf{r}_i(t) - \mathbf{r}_i(0)$ is the displacement vector of particle $i$ over time period $t$ starting in a specified MD cell at $t=0$. A different choice of $t=0$ supercell will lead to a different value of polarization at time $t$.    

Eq.~\ref{piter} is not new in classical MD. Polarization defined as a time integral of current is familiar from MD studies of homogeneous electrolytes and was referred to as itinerant polarization~\cite{Caillol:1989jcpa,Caillol:1989jcpb,Caillol:1994ho}. While itinerant polarization has the form of a dipole moment density, it is fundamentally different from the  cell polarization  obtained by evaluating Eq.~\ref{defpz} for point charges
\begin{equation}
\label{pcellspc}
 \mathbf{P}^{\textrm{cell}}(t) =  \frac{1}{\mathit{\Omega}} 
\sum_i^{\textrm{cell}} q_i \textrm{nint}(\mathbf{L}^{-1}\mathbf{r}_i(t))
\end{equation}
where $\mathbf{L}$ is the supercell matrix and $\textrm{nint}(x)$ is the nearest-integer rounding function. This equation is the 3D particle form of Eq.~\ref{defpz}.

The difference between Eq.~\ref{piter} and \ref{pcellspc} is that itinerant polarization preserves the continuity of time integrated current. The particles must be followed if they leave the MD cell where they started out at $t=0$. This is how the self diffusion coefficient is calculated from the mean square displacement. In contrast, cell polarization is calculated from the particles in the MD cell at all times. If a particle crosses the boundary of the cell it is replaced by its periodic image entering from the opposite side. 
  
The history of itinerant polarization in classical simulation of ionic solutions goes beyond the issue of calculating transport coefficients touched upon above\cite{Caillol:1989jcpa,Caillol:1989jcpb,Caillol:1994ho}. The problem arises when a surface term is included in Ewald summation\cite{DeLeeuw:1980rs80a,DeLeeuw:1980rs80b,Ballenegger:2014jpc}. These surface terms represent the effect of embedding the system in an  environment other than a metal (for example vacuum). Ewald surface terms are proportional to the square of polarization, a feature they have in common with the Vanderbilt constant $\bar{D}$ hamiltonian\cite{Zhang:2015ms, Zhang:2016ho}(see also the discussion in section \ref{sec:dpzc}).  Whenever an ion crosses a cell boundary, the cell polarization jumps and hence the forces.  To avoid these discontinuities, the polarization was set equal to the itinerant polarization. However, the problems with cell polarization go deeper than this seemingly technical inconvenience for MD. In his 1994 paper, Caillol showed, after what he called a long struggle with the subject, that cell polarization defined in Eq.~\ref{pcellspc} violates fundamental relations for ionic solutions, such as the Stillinger-Lovett sum rule \cite{Caillol:1994ho}. This means that the fluctuations are incorrect. There are no such difficulties with itinerant polarization, which satisfies all key statistical mechanical conditions.    
   
 \subsection{Polarization and electric displacement at ZNC}
\label{sec:dpzc}
Working through the analytical dielectrics of the continuum model we found that the charge of the EDL's could only be balanced for a finite macroscopic field $\bar{E} = \bar{E}_{\textrm{znc}}$ given in Eq.~\ref{pzc}. This field was related to the polarization at ZNC in Eq.~\ref{polpzc} which states that at ZNC the macroscopic field is identical to the polarization field ($\bar{E}_{\textrm{znc}} = - 4 \pi P^{\textrm{cell}}_{\textrm{znc}}$). Substitution in Eq.~\ref{defdbar} leads to the conclusion that $D_{\textrm{znc}} = 0$. Evidently ZNC corresponds to open circuit conditions with a fully unscreened depolarizing field.  Cell boundaries in the vacuum implies that the cell polarization equals the itinerant polarization. Hence for the ECS geometry (Fig.~\ref{ecs}),  we can safely assume that the predictions of the continuum model  carry over to the ECS atomistic model, suggesting that we set $\bar{D}=0$ in Eq.~\ref{uvdb}. The SSV internal energy Hamiltonian  is simplified to 
\begin{equation}
\label{depofield}
H_D\left( v ,\bar{D}=0 \right) = H_{\mathrm{PBC}}(v)+  2\pi\Omega P^2 
\end{equation}
The polarization coupling term is the familiar YB/dipole correction~\cite{Neugebauer:1992uh,Bengtsson:1999il, Yeh:1999dm}, which has been shown to be a successful and computationally efficient scheme for eliminating ``spurious''  electrostatic interactions in a periodic slab-vacuum geometry. As shown here, SSV theory gives this popular and simple dipole correction scheme a  dielectric and thermodynamic foundation.  

The problem addressed in this paper is how to remove the finite size effect of a supercell with boundaries in the electrolyte (ICS, Fig.~\ref{ics}). Such a scheme would also allow us to simulate charged insulator-electrolyte interfaces without introducing an artificial vacuum slabs.  The ICS and ECS  are representations of the same periodic system and should ultimately give consistent results. This was verified for the continuum model in sections \ref{sec:sternsc} and \ref{sec:sternsc_D}. The equations of state for the polarization, different at first, could be reconciled  when the multivaluedness of polarization admitted by the modern theory of polarization was taken into account\cite{King-Smith:1993prb,Resta:1994rmp,Resta:2007ch}. This argument will now be transferred and generalized for the atomistic system.  

The solvent polarization is evaluated following the familiar scheme of adding up the molecular dipoles. While this definition of solvent polarization is often viewed as an approximation to the point dipoles of the physical chemical theory of polar molecular liquids, from a formal perspective it can be seen as itinerant polarization conforming to Eq.~\ref{piter} (see also Ref.~\citenum{Zhang:2015ms}). The multivaluedness of the itinerant dipole is therefore entirely due to the contribution of the ions. The difference can only be a multiple of $+e L$ or $-e L$, as the ions are monovalent and the plate charge is always an integer multiplied by $e$ in our system. The sign depends on whether the ICS is shifted to the left or right with respect to the ECS. To obtain the corresponding itinerant polarization, we divide by $LA$. Polarization in the ICS and ECS for the atomistic model is thus related according to
\begin{equation}
\label{Pmulti}
P_{\textrm{ICS}}^{(n)} = P_{\textrm{ECS}} + n\frac{e}{A}
\end{equation}
where $n$ is a positive or negative  integer (or zero).  $e/A$ is the quantum of polarization\cite{Resta:2007ch}.  

The set of itinerant polarization values  for given $n$ is called a branch. Itinerant polarization follows the branch determined by the initial configuration at $t=0$.  All branches are a representation of the same physical state. The dynamics of the atoms can therefore not differ if we change branch. The dynamics is driven by the electric field, which should therefore be also the same on every branch. Hence, when the polarization is changing with the choice of supercell, so must the electric displacement in order to conserve the electric field  
\begin{equation}
\label{Dmulti}
\bar{D}_{\textrm{ICS}}^{(n)} = \bar{D}_{\textrm{ECS}} + n\frac{4 \pi e}{A}
\end{equation}
 Eq.~\ref{pics2ecs} and \ref{dics2ecs}  are a special case of Eq.~\ref{Pmulti} and \ref{Dmulti}. The shift due to a change of supercell in the continuum model is entirely determined by the (integer) plate charge $\sigma_0$. 

The reference polarization or electric displacement in Eqs.~\ref{Pmulti} and \ref{Dmulti} is the value for an ECS, which is unique (the only case when cell polarization equals itinerant polarization). As we have seen, ZNC corresponds to a state with $\bar{D}_{\textrm{ECS}}=0$. Insertion in Eq.~\ref{Dmulti} yields for the ICS displacement field at ZNC:  
 \begin{equation}
\label{D_pznc_multi}
D_{\textrm{znc}}^{(n)} =  n\frac{4\pi e}{A}
\end{equation}
We have omitted the ICS label. The label has  become redundant because according to Eq.~\ref{D_pznc_multi} the displacement field at ZNC is a multiple of the polarization quantum, including $n=0$ which covers the ECS state.   

Summarizing, what we have achieved is a generalization of the YB/dipole correction method for a  periodic MD cell. Now the supercell can be either electrolyte centred surrounded by vacuum (Fig.~\ref{ecs}), as in the original YB/dipole correction method~\cite{Neugebauer:1992uh,Bengtsson:1999il, Yeh:1999dm}, or insulator centred (Fig.~\ref{ics}). In the latter case we may have to impose a finite value of electric displacement (Eq.~\ref{D_pznc_multi}).  If so, the appropriate Hamiltonian  to use is the SSV constant-D Hamiltonian of Eq.~\ref{uvdb} with itinerant polarization (Eq.~\ref{piter}) instead of cell polarization (Eq.~\ref{pcellspc}). 

We are now free to eliminate the vacuum and fill it with a material insulator instead. The location of the supercell boundary should not matter provided the correct definition of polarization is used.  The vacuum gap is (in classical simulation) a convenient but unnecessary device.  The argument that led to this generalized open circuit scheme is based on abstract concepts of the modern theory of polarization originally developed for electronic systems. The scheme will have to be validated  by atomistic simulations as reported in the next Section.

\section{Results} \label{sec:results}

\subsection{Model system and molecular dynamics} \label{sec:spcmd}

The periodic EDL model system used to validate the finite field methods presented in the method sections is the classical SPC system of Figs.~\ref{ecs} and \ref{ics}. The electrolyte consists of 202 water molecules, 5 Na$^+$ and 5 Cl$^-$ ions. The polar insulator slab is modelled as two rigid uniformly charged atomic walls separated by a vacuum space. The simulation box is rectangular. The length in $x$ and $y$ direction is 12.75~\AA~ and the length in $z$ direction varies from 61.24~\AA~to 112.48~\AA~depending on the thickness of the vacuum layer acting as the insulator. 

Water is described by the SPC/E model potential~\cite{Berendsen:1987uu} and kept rigid using the SETTLE algorithm~\cite{Miyamoto1992}. Na$^+$ and Cl$^-$ ions are modelled as point charge plus Lennard-Jones potential using the parameters from Jung-Cheatham~\cite{Joung2008}. This set of parameters is known to reproduce the correct excess chemical potential of an aqueous  NaCl solution at finite ionic strength~\cite{Zhang:2010zh,Zhang:2012fo}. The Van der Waals parameters of the interaction sites of the rigid wall are simply chosen identical to those of water oxygen atoms. The MD integration time step is 2~fs. Unless mentioned elsewhere, each MD trajectory is collected for 1ns. The Ewald summation is implemented using the Particle Mesh Ewald (PME) scheme~\cite{Ewald}. Short-range cutoffs for the Van der Waals and Coulomb interaction in direct space are 6~\AA. The temperature is controlled by a Nos\'e-Hoover chain thermostat set at 298K~\cite{martyna92} . All simulations are carried out with a modified version of GROMACS 4 package~\cite{Hess2008}. 

One needs to pay attention that when computing the itinerant polarization according to Eq.~\ref{piter}, to use the same $\mathbf{P}(0)$ as the reference for consistent results.

\subsection{ZNC and Helmholtz capacitance at constant electric field $\bar{E}$}
\label{sec:Eznc}

\begin{figure} [h]
\includegraphics[width=0.95\columnwidth]{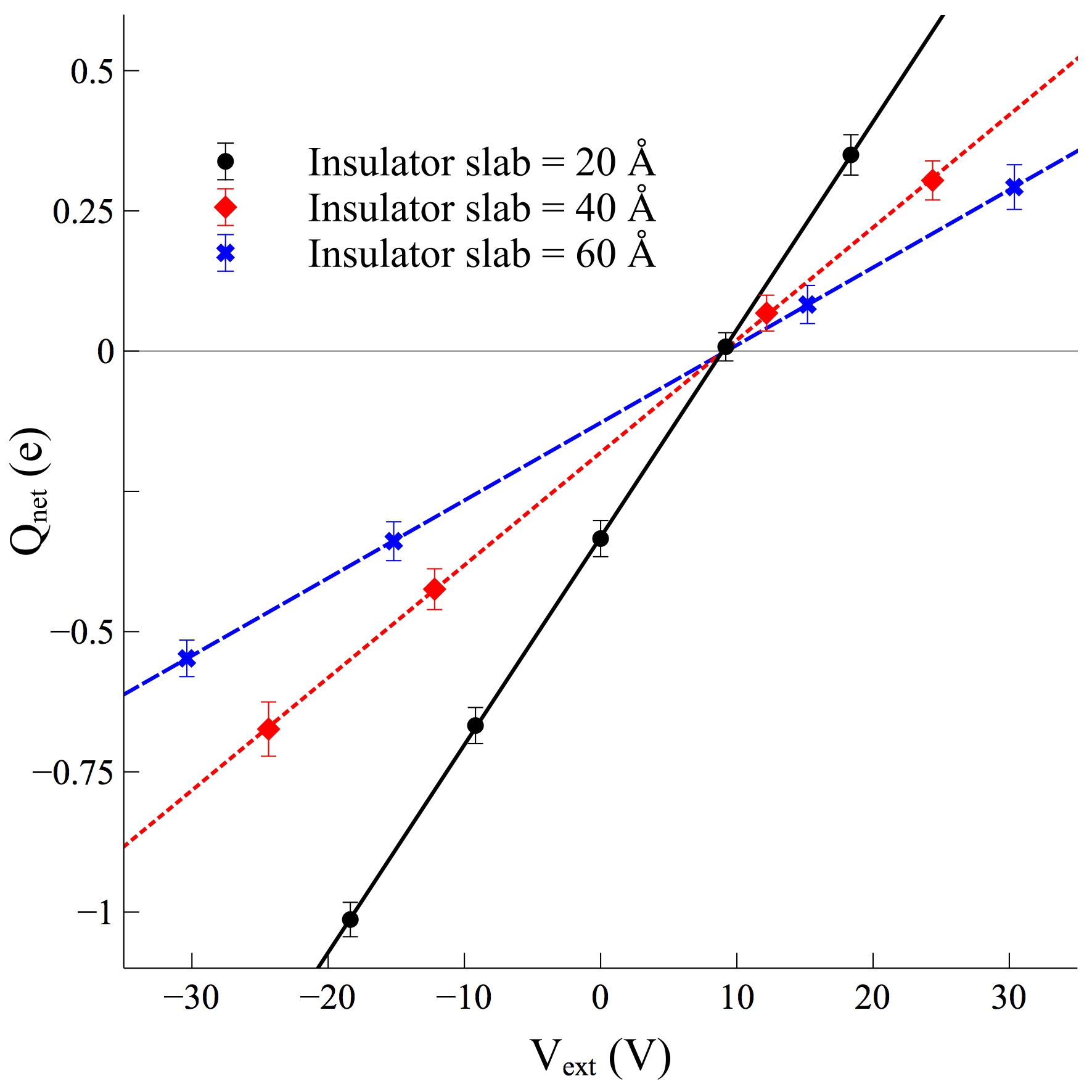}% Here is how to import EPS art
\caption{\label{QnetV} The net EDL charge $Q_\text{net} = A(-\sigma_0 + \sigma)$ as a function of the external voltage $V_{\text{ext}} = - L \bar{E}$ ($\bar{E}$ is the constant macroscopic field in Eq.~\ref{fhmd}). Plotted are the results for three values of thickness of insulator slabs for the ECS system of Fig.~\ref{ecs}. The full plate charge is $2.0 e$. The lines intersect at zero net charge (ZNC) $Q_\text{net}=0$. The potential at ZNC directly  gives the Helmholtz capacitance $C_\text{H}$ according to Eq.~\ref{pzc} as derived for the continuum model. }
\end{figure}

In this section the Helmholtz capacitance $C_{\text H}$ of the EDL is estimated from the macroscopic electric field at ZNC using Eq.~\ref{pzc}.  The point of ZNC is located by varying the electric field $\bar{E}$ in the electrolyte centred supercell Fig.~\ref{ecs} and computing the net charge $Q_{\textrm{net}}$.  The charge density on an insulator surface is set to a fixed  value  $\sigma_0$ amounting to a total charge of $\sigma_0 A = 2e$. The charge distribution in the electrolyte is obtained from a MD trajectory generated by the  SSV constant-E Hamiltonian of  Eq.~\ref{fhmd}.  $Q_{\textrm{net}}$  is computed according to the method explained in section \ref{sec:edlch}. The variation of charge with the length of the insulator slab at $\bar{E}=0$ (standard Ewald) was presented in Fig.~\ref{Qz}.  $Q_{\textrm{net}}$ is again plotted in Fig.~\ref{QnetV} as a function of $V_{\textrm{ext}}=-L\bar{E}$ at selected values of the size of the insulator slab ($l_d$ in section \ref{sec:polstern}). The potential $V_{\textrm{ext}}$ at ZNC is an intensive property independent of system size an therefore a more suitable quantity to represent our results than the field $\bar{E}$. 

 As can be seen from Fig.~\ref{QnetV},  $Q_\text{net}$ increases linearly with $V_\text{ext}$. Lines for different insulator slabs intersect at one point where $Q_\text{net} = 0$. These observations indeed follow Eq.~\ref{Qnet_stern_E} derived by solving the highly simplified Stern-like continuum model. The external voltage at ZNC is $L\bar{E}_{\textrm{znc}} =8.9$ V.  With a plate charge of $\sigma_0 A = 2e$ this leads to an overall capacitance of 3.2~\AA~corresponding to $C_\textrm{H}/2 = 2.2 \, \mu$F/cm$^2$. Assuming that the two EDLs are equivalent, the capacitance of each EDL is 4.4 $\mu$F/cm$^2$. 
 
\begin{figure} [h]
\includegraphics[width=0.95\columnwidth]{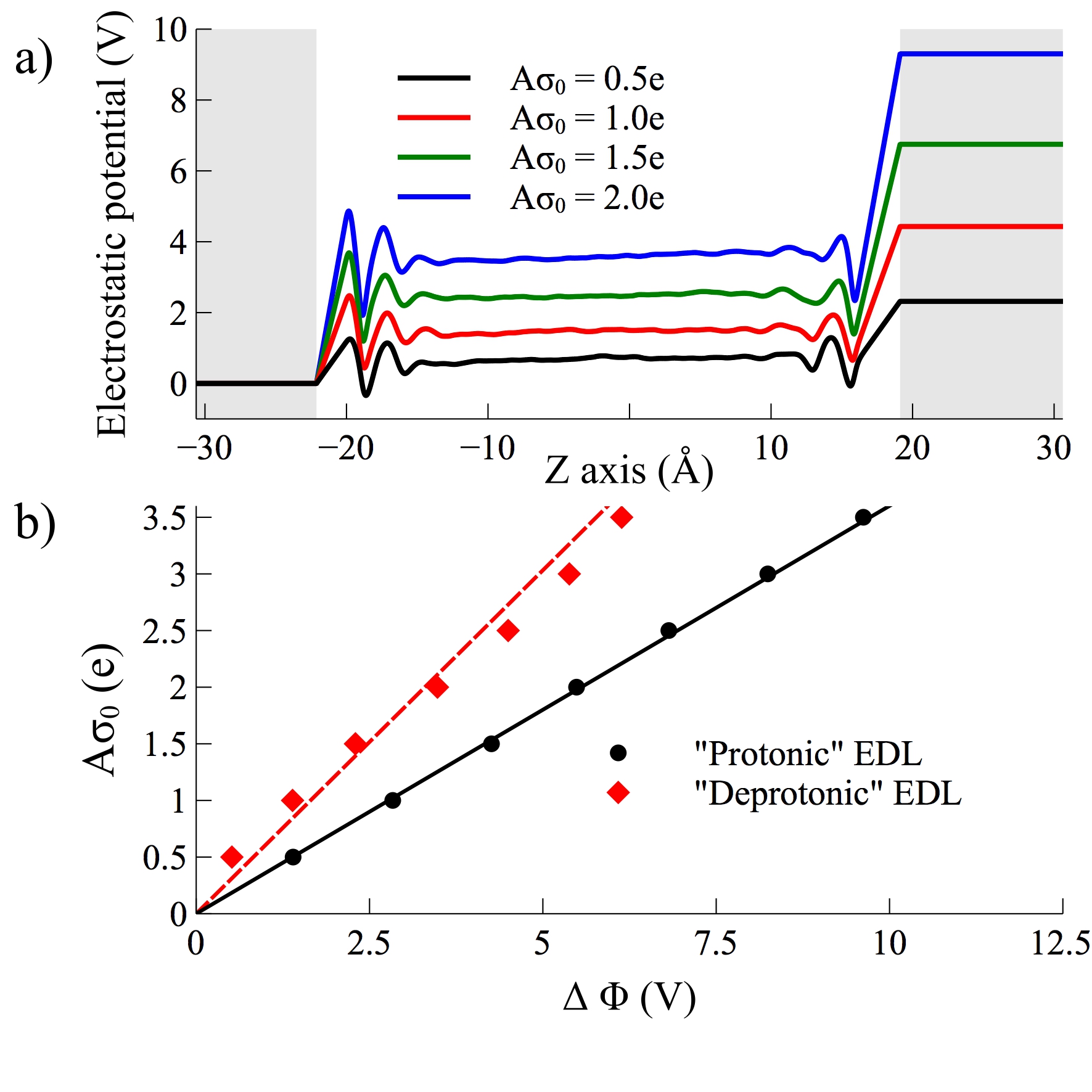}% Here is how to import EPS art
\caption{\label{dipole_corr} a) The electrostatic potential profile of the ECS  of Fig.~\ref{ecs} for different surface charge densities $\sigma_0$. $\bar{D}=0$ electrostatic boundary (corresponding to the YB/dipole correction) are applied; b) Surface charge as function of the computed potential drop $\Delta \Phi$ over the interface. ``Protonic'' refers to the EDL at the positively charged insulator surface, ``Deprotonic'' refers to the EDL at the negatively charged surface.}
\end{figure}

The field estimate of capacitance was compared to the value computed  using the conventional method based on changes in the electrostatic interface potential $\Delta \Phi$ in combination with the YB/dipole correction scheme~\cite{Neugebauer:1992uh,Bengtsson:1999il, Yeh:1999dm}, i.e. $\bar{D}=0$.  As explained,  $\bar{D}=0$ electric boundary condition impose ZNC on the ECS forcing the field in the insulator ($E_d$ in Eq.~\ref{Edc}) to vanish. The electrostatic potential is constant  in the insulator slab and $\Delta \Phi$ is simply the difference with the flat part of the electrostatic potential in the electrolyte (Fig.~\ref{dipole_corr}a).  This method gives a separate capacitance for the positively (``protonic'') and negatively (``deprotonic'') charged surface. We calculated 3.5 $\mu$F/cm$^2$ respectively 5.5 $\mu$F/cm$^2$ (Fig.~\ref{dipole_corr}b). To compare to the field estimate of the overall capacitance $C_{\textrm{H}}/2 = 2.2 \mu$F/cm$^2$, we must take the series total capacitance. This gives $2.1\mu$F/cm$^2$ which is in excellent agreement. This value, although rather low, is in the range of capacitance measured and computed for metal electrodes~\cite{Schmickler:2010Sc}. Note however, that the potential of an metal electrode is constant. The charge density fluctuates in that case, while it is fixed on the surface of the insulator in our model (for an in-depth discussion see Merlet et al.~\cite{Merlet:jpcl13} and Limmer et al~\cite{Limmer:prl13}).   

\begin{figure}[h]
\includegraphics[width=0.95\columnwidth]{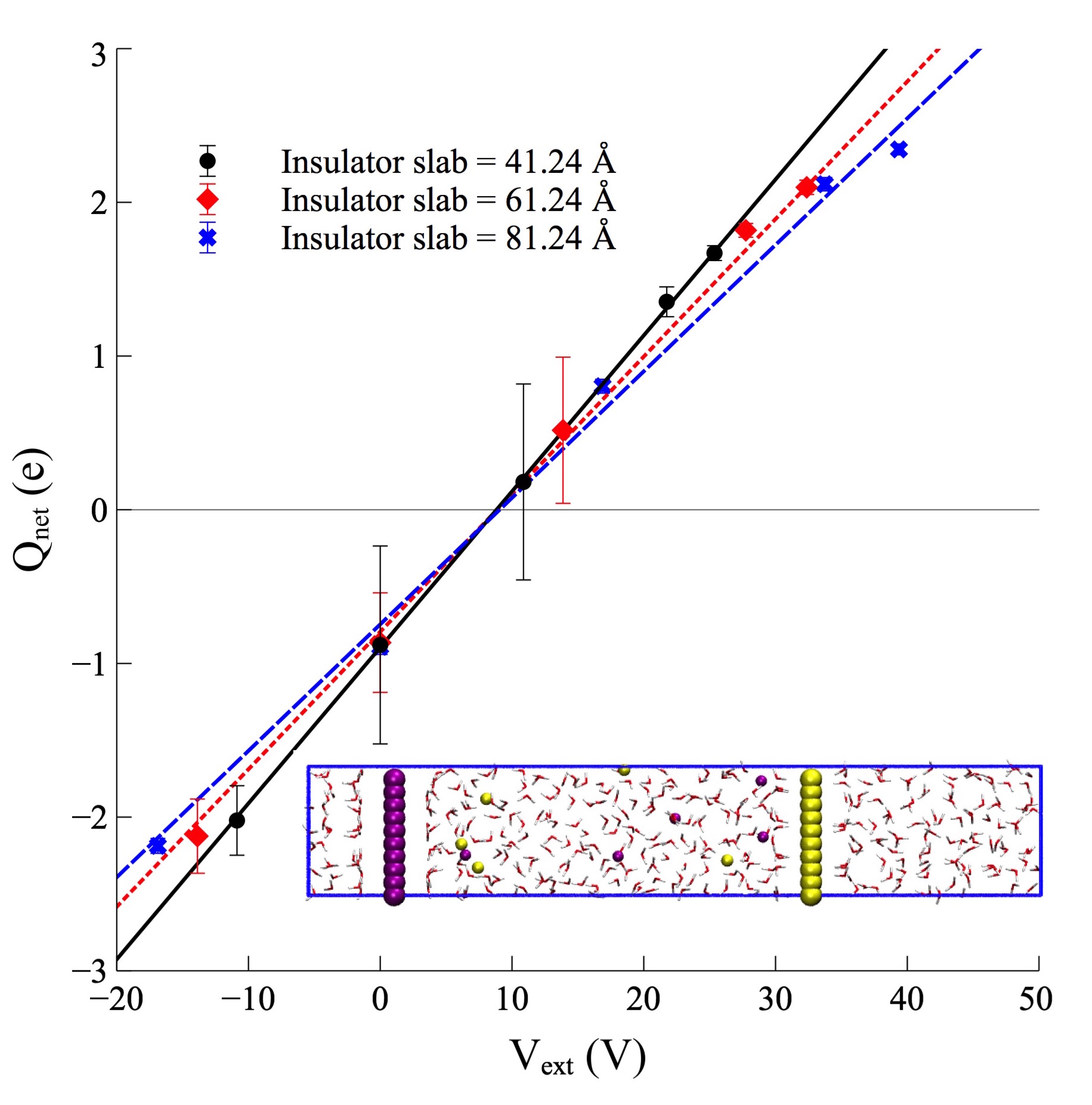}% Here is how to import EPS art
\caption{\label{waterins} Net EDL charge $Q_\text{net}$  as function of external voltage $V_{\text{ext}}$ with liquid water filling the insulator space (see inset). The material insulator changes the value of the slopes for different thickness of insulator slabs  (compare Fig.~\ref{QnetV}) but has negligible effect on the location of the ZNC intersection point.  }
\end{figure}

 To verify that the method also works for more realistic insulator slabs, we replaced the vacuum with liquid SPC/E water (Fig.~\ref{waterins} inset). Apart from this change, all model and simulation settings are the same as used for producing Fig.~\ref{QnetV}. The resulting $Q_\textrm{net}$ dependence on external voltage  $V_{\text{ext}}$ is shown in Fig.~\ref{waterins}. Consulting the relevant equation for the continuum model, Eq.~\ref{Qnet_stern}, enhancement of the dielectric constant $\epsilon_d$ of the insulator should lead to a larger slope of the $Q_\text{net} (V_\text{ext})$ line. This is indeed what we find. The common intersection point is slightly shifted away from $Q_\text{net}=0$ because of a more scattered distribution of the data. At $Q_\text{net}=0$, we find $V_\text{ext}$ is 8.3 V, yielding an overall capacitance $C_\text{H}/2$ of $3.0$\AA \, or 2.1 $\mu$F/cm$^2$. These values are very close to what we found in Fig.~\ref{QnetV} without water. Evidently, the dielectric properties of the insulator have negligible effect on the interface capacitance in our SPC model. Thus, for the sake of simplicity, we stayed with the vacuum insulator model shown in Fig.~\ref{ecs} and Fig.~\ref{ics} for further tests.

 \subsection{Electric equations of state:  $ P (\bar{E})$ and $P(\bar{D})$}
\label{sec:eosmd} 

The point of zero net charge of the periodic model system is the physical relevant state for  modelling of an interface between a charged surface of a semi-infinite insulator and an electrolyte.  For  the periodic system.  the ZNC  state is just a single  point of the electric EOS.  Changes in polarization $P$ in response to variation in $\bar{E}$ and $\bar{D}$ away from ZNC are strongly conditioned by the periodicity and finite size of the model system. However, they  are also of interest, if only for an analysis of the alternative method of computing capacitance using the derivative of polarization wrt to fixed surface charge (Eq.~\ref{deltaP_D}). This is the subject of the present section. 

Again, we will exploit the special properties of an electrolyte centred supercell (ECS) with a vacuum gap for insulator (Fig.~\ref{ecs}). For the ECS geometry the cell polarization equals the itinerant polarization. We take these unambiguous results as our benchmark for the more involved (and potentially confusing) insulator centred cell (ICS) calculations which should ultimately produce the same value of capacitance.  We start by checking whether the expectation value $\langle P \rangle$ from the all-atom simulation agrees with the continuum EOS of Eqs.~\ref{eosE} and \ref{eosD}. To facilitate the comparison, $\langle P\rangle$, $\bar{E}$ and $\bar{D}$ are given in the same Gaussian units of surface charge density (C/m$^2$). 

\begin{figure} [h]
\includegraphics[width=0.95\columnwidth]{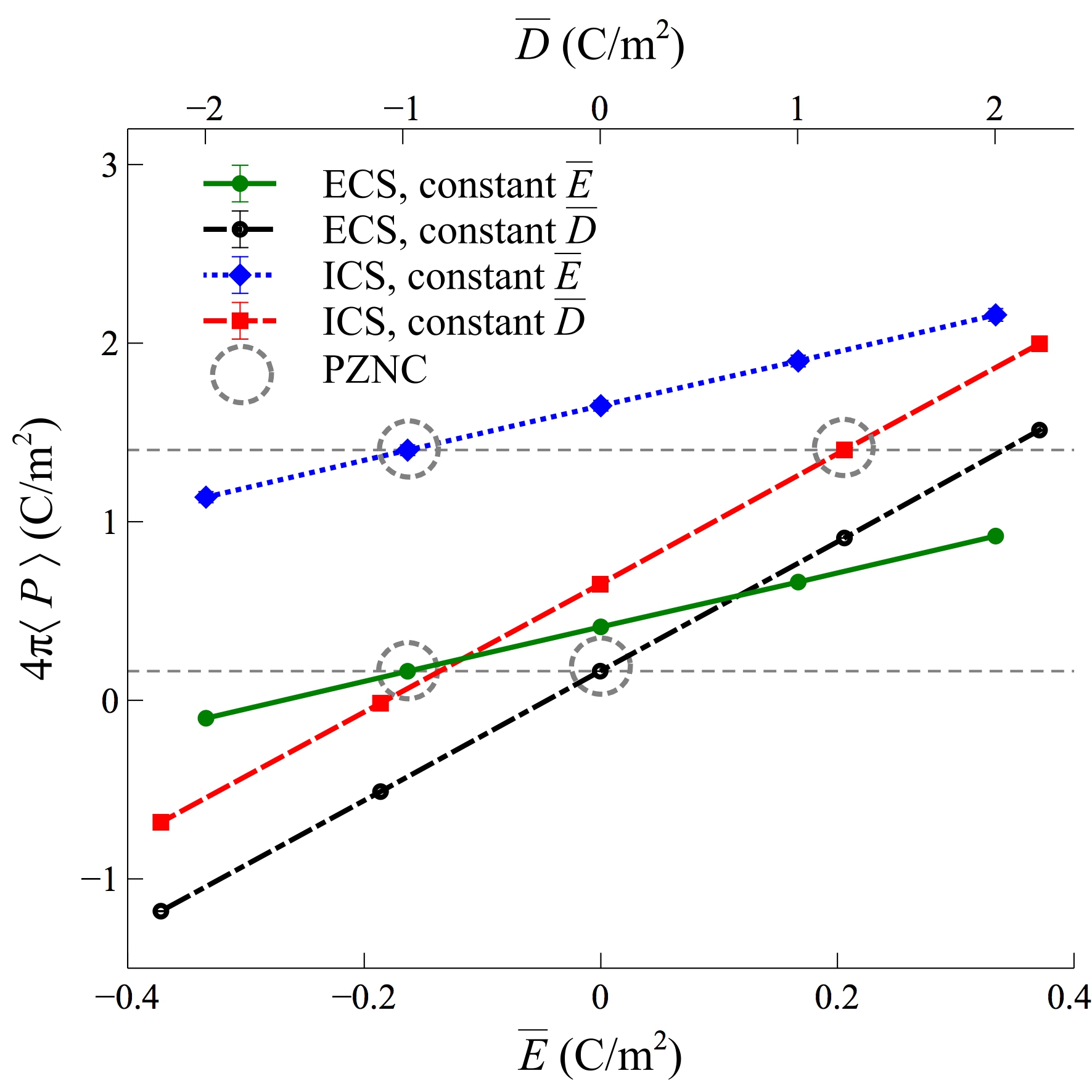}
\caption{\label{PDEeos} Computed electric equations of state $P (\bar{E})$ and $P(\bar{D}) $ for the charged insulator-electrolyte interface in ECS and ICS periodic boundary conditions. The supercell dimensions are the same as in Figs.~\ref{ecs} and \ref{ics}. In particular the critical width of the vacuum layer $l_{\textrm{d}}=20$ \AA. The point of zero net charge (PNZC) is indicated by circles. Analytical equations of state as derived for the continuum model are given by Eqs.~\ref{eosE} and \ref{eosD} for the ECS and Eqs.~\ref{ICSeosE} and \ref{ICSeosD}  for the ICS. The relative shift of parallel lines is due to a difference in quantum of polarization (see text). }
\end{figure}

The result for ECS polarization as function of macroscopic field is the green line shown in Fig.~\ref{PDEeos}. $\langle P_{\textrm{ECS}}\rangle$ increases linearly with $\bar{E}$ in accordance with Eq.~\ref{eosE}. The finite polarization induced by the fixed surface charge $\sigma_0$ is in turn screened by the electrolyte. This should be reflected in a ``global'' dielectric permittivity $\bar{\epsilon}$  of Eq.~\ref{bareps} larger than unity and therefore a positive slope just as the simulation result indicates. At ZNC (obtained by monitoring $Q_{\textrm{net}}$), $4\pi \langle P_{\textrm{ECS}}\rangle = 0.1626$ C/m$^2$ and $\bar{E} = -0.1633$ C/m$^2$. Therefore the relation $4\pi P_{\textrm{znc}} = -\bar{E}_{\textrm{znc}}$ (Eq.~\ref{polpzc}) holds in the atomistic system.  

The EOS for the ICS is obtained from the very same MD trajectory. Only the evaluation  of polarization is different. $\langle P_{\textrm{ICS}}(\bar{E})\rangle$ is shown as the blue dotted line in Fig.~\ref{PDEeos}.  The ICS line runs parallel to the ECS line, but shifted. The shift  is $1.24$ C/m$^2$ agreeing to high accuracy with $4 \pi$ times  the quantum of polarization for our model. Therefore, the MD EOS satisfy Eq.~\ref{Pmulti} for $n=1$.  We also verified that evaluating the polarization using  expression Eq.~\ref{pcellspc} for cell polarization rather than Eq.~\ref{piter} for itinerant polarization reproduces Eq.~\ref{pcellics}. The computed slope of $4 \pi P^{\textrm{cell}}_{\textrm{ICS}}$ vs $\bar{E}$ (not shown) is indeed $-1$ and inconsistent with the slope of ECS polarization. In contrast, for itinerant polarization the ICS and ECS slopes match. The atomistic simulation therefore confirms  the theoretical analysis of the continuum model. Cell polarization violates fundamental relations for ionic solutions. 

Fig.~\ref{PDEeos} looks somewhat cluttered, but we thought it instructive to plot $P(\bar{E})$ and $P(\bar{D})$ in the same figure. The result for $\langle P\rangle$ as a function of $\bar{D}$ for the ECS is shown as the black dash-dotted line. Similar to $\langle P_{\textrm{ECS}}(\bar{E}) \rangle$, $\langle P_{\textrm{ECS}}(\bar{D}) \rangle $ increases linearly with $\bar{D}$, however with a smaller slope. This behaviour is in agreement with the continuum model. According to  Eqs.~\ref{eosE} and \ref{eosD} the slope of $P$ vs $\bar{E}$ exceeds the slope vs $\bar{D}$ by a factor $(\bar{\epsilon}-1)/(1-1/\bar{\epsilon})=\bar{\epsilon}$. The value of $\bar{\epsilon}$ is a measure of the net dielectric screening of a uniform applied electric field by the heterogeneous system. It is a rather modest effect in our SPC model.  Fitting to the derivatives of $\langle P(\bar{E}) \rangle $ and $\langle P(\bar{D}) \rangle$ gives $\bar{\epsilon} =2.524$ respectively 2.528. At ZNC, $\bar{D}=0$ as expected from the discussion in Section~\ref{sec:dpzc}. The ZNC point for the $P_{\textrm{ICS}}$ is however not at $\bar{D}=0$.  By monitoring $Q_{\textrm{net}}$, it is found at $\bar{D}=1.237$ C/m$^2$,  effectively equal to $4\pi$ times the quantum of polarization, i.e. $4\pi e/A=1.238$ C/m$^2$. Our result for the electric displacement at ZNC  is therefore an example of  Eq.~\ref{D_pznc_multi} for $n=1$. 

What remains to be explained is the shift between the ICS and ECS $P(\bar{D})$ lines in Fig.~\ref{PDEeos}. The gap at given value of $\bar{D}$ is not $4\pi$ times the polarization quantum $e/A$, in contrast to what we found for $P(\bar{E})$. This is the effect anticipated in our analysis of the $P_{\textrm{ICS}}(\bar{D}_{\textrm{ICS}})$ equation of state (Eq.~\ref{ICSeosD}) at the end of section \ref{sec:sternsc_D}. According to this ``corresponding states'' argument it should be possible to reconstruct the black dash-dot line from the red dashed line. It is a two-step procedure: First down-shift the red dashed line by $4\pi e/A$ and then left-translate it by $4\pi e/A$. This leads to overlap of the point of ZNC in the red dashed and black dash-dot line.  Applying the same procedure to Eq.~\ref{ICSeosD}, one finds that the shift between branches at constant $\bar{D}$ is $(4\pi e/A) \bar{\epsilon}^{-1}$. Since $\bar{\epsilon} > 1 $, the gap at constant $\bar{D}$ is smaller than $4\pi$ times the polarization quantum, as seen in Fig.~\ref{PDEeos}.

\subsection{Capacitance from polarization increments at constant electric displacement $\bar{D}$}
\label{sec:deltaP}
We now take the second route to calculate the Helmholtz capacitance using the increment in the polarization at constant electric displacement $\bar{D}$. The equation to use is Eq.~\ref{deltaP_D} as obtained for the continuum model.  Eq.~\ref{deltaP_D} is derived from the first term of the $P(\bar{D})$ continuum equation of state Eq.~\ref{eosD}.  The second term, specifying the linear variation with $\bar{D}$ was found to be transferable in Section \ref{sec:eosmd} validating the variation with the shifted displacement field as specified by Eq.~\ref{ICSeosD}. Therefore, we can expect with some confidence that Eq.~\ref{deltaP_D} remains valid for the ECS atomistic system. But the question is whether this relation is transferable to itinerant polarization in the ICS atomistic system.

\begin{figure} [h]
\includegraphics[width=0.95\columnwidth]{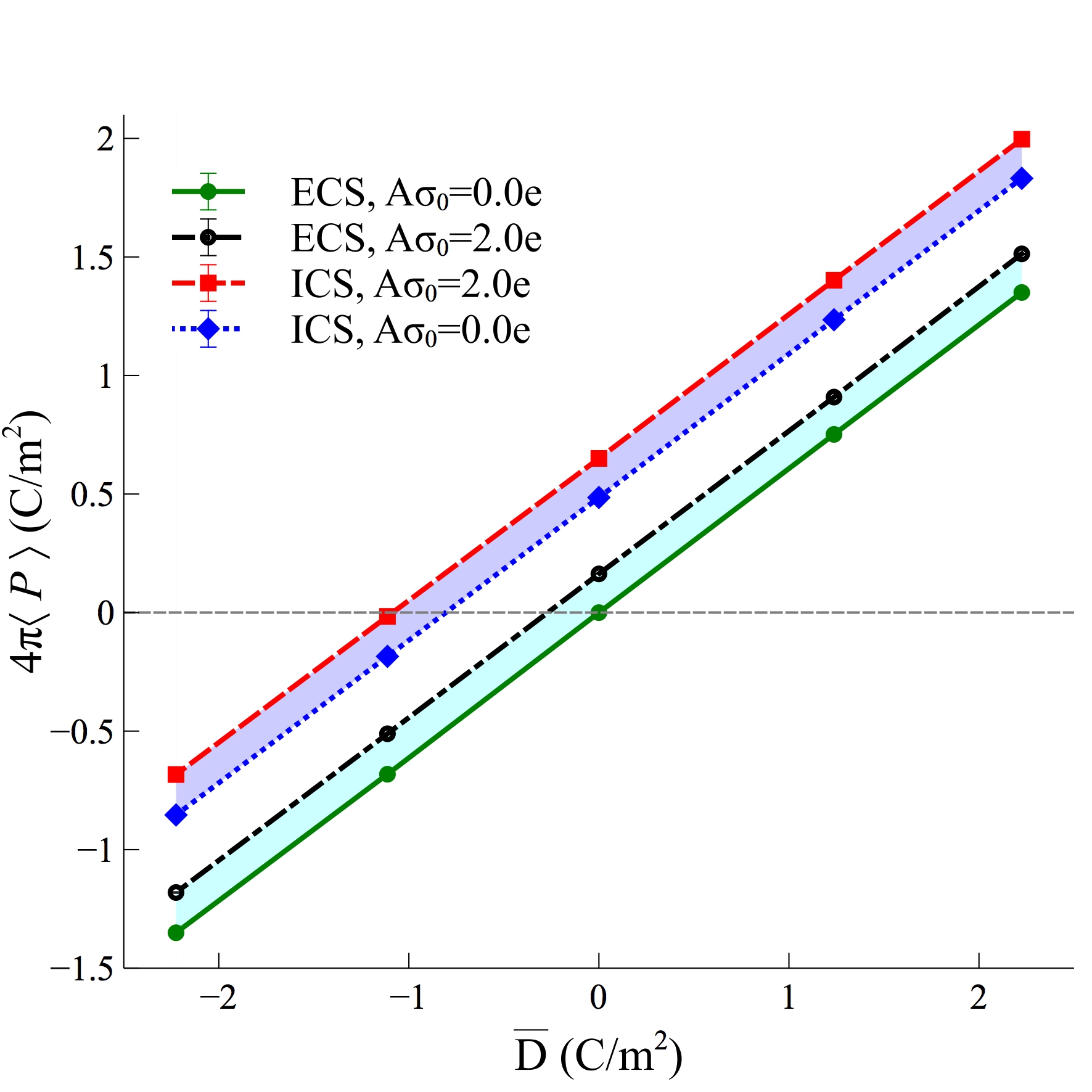}% Here is how to import EPS art
\caption{\label{deltaP} Difference in polarization $P(\bar{D})$ at plate charges $A\sigma_0=2.0e$ and $A\sigma_0=0.0e$ shown as coloured bands for the electrolyte centred supercell (ECS, Fig.~\ref{ecs} ) and insulator centred supercell (ICS, Fig.~\ref{ics}). Applying Eq.~\ref{deltaP_D} the increment of polarization at constant $\bar{D}$ is used to estimate the EDL capacitance.}
\end{figure}

The derivative of $P(\bar{D})$ with respect to $\sigma_0$ is estimated from finite differences comparing the polarization at the plate charge density  $A\sigma_0=2.0e$ and $A\sigma_0=0.0e$. The results are shown in Fig.~\ref{deltaP}. To relate the ICS and ECS data we apply the corresponding state transformation outlined in section \ref{sec:eosmd}.  Analyzing  Fig.~\ref{PDEeos} we found that for the  $A\sigma_0=2.0e$ system the  ICS and ECS displacement fields at the same state point differ by $4\pi e/A$.  Furthermore, the polarization for the  ICS at $A\sigma_0=0.0e$ shown in Fig.~\ref{deltaP} has been moved down by  $2.0e/A$ to make a proper comparison with that for ICS at $A\sigma_0=2.0e$. This adjustment is the correction for the polarisation quantum due to the plate charge $A\sigma_0=2.0e$  as given by Eq.~\ref{pics2ecs}.  

The increment  $ \langle \Delta P \rangle$ for a given supercell is highlighted in Fig.~\ref{deltaP} by coloured bands. The width of the bands is constant consistent with the linearity of Eqs.~\ref{eosD} and Eq.~\ref{ICSeosD}.   $\langle \Delta P \rangle$  remains the same throughout the full range of $\bar{D}$ for both choices of the supercell. The corresponding Helmholtz capacitance $C_{\textrm{H}}$ is 4.5 $\mu$F/cm$^2$ for the ECS and 4.4 $\mu$F/cm$^2$ for the ICS. These numbers are in excellent agreement with the estimate calculated from $V_{\textrm{ext}}$ at ZNC in section \ref{sec:Eznc}.  Therefore, Eq.~\ref{deltaP_D} is validated for both ECS and ICS atomistic systems.

How is the estimate of capacitance obtained from the polarization related to the standard approach of varying the plate charge and computing the changes  in the electrostatic potential? We did the calculation  (Fig.~\ref{dipole_corr}) and the results were already reported in section \ref{sec:Eznc}.  The MD simulation was carried out in an ECS using the  YB/dipole correction~\cite{Yeh:1999dm, Neugebauer:1992uh,Bengtsson:1999il} which correspond to $\bar{D}=0$ and, hence, $\langle \Delta V \rangle = -L\langle \bar{E} \rangle = 4\pi L\langle \Delta P\rangle$. Therefore, computing the difference in the polarization at $\bar{D}=0$ is equivalent to determining $\partial \sigma_0 /\partial \langle \Delta V \rangle $. However because of the apparent absence of non-linear effects, the polarization increment can be calculated at any value of $\bar{D}$. It is not necessary to search for the PZNC. In fact, Eq.~\ref{pzc2} allows us to turn the argument around and determine the electric field $\bar{E}_{\textrm{znc}}$ at ZNC from the polarization charge derivative at constant $\bar{D}$. This gives an $\bar{E}_{\textrm{znc}}$  of $-0.165$ C/m$^2$  for the ECS and $-0.167$ C/m$^2$ for the ICS. These estimates are in good agreement with $\bar{E}_{\textrm{znc}}=-0.1633$ C/m$^2$ obtained from the $Q_{\textrm{net}}=0$ field in $\bar{E}$ controlled simulations (section \ref{sec:Eznc}) . Therefore, constant $\bar{D}$ simulations also provide a convenient alternative for locating the point of ZNC without having to compute the net charge $Q_{\textrm{net}}$. This is a definite advantage in electronic structure calculation.

\subsection{Convergence of the capacitance calculations}

The virtually  noiseless data shown in Figs.~\ref{QnetV}, \ref{PDEeos} and \ref{deltaP} are the result of  averaging over nanosecond MD trajectories.  MD runs of this length are routine for SPC force field MD simulation but still not yet feasible for DFT based MD simulation of a similar size system. Therefore, in this section, we investigate the convergence with time of capacitance calculated with the field methods proposed here. As we saw in Fig.~\ref{deltaP}, the average of $\Delta P$ (therefore $C_{\textrm{H}}$) is not sensitive to the value of  $\bar{D}$  imposed as electric boundary condition.  This is not necessarily the case for the convergence. The time scale of the fluctuations could be different. For an  example of the effect of electrostatic boundary conditions on polarization dynamics we refer to our previous study of Ref~\citenum{Zhang:2015ms}.  This concerned a comparison of the fluctuations at constant $\bar{E}$  and constant $\bar{D}$ in pure liquid SPC water.   The relaxation time of polarization in the constant $\bar{D}$ ensemble is more than an order of magnitude shorter than in the constant $\bar{E}$ ensemble. This corresponds to the difference between longitudinal and transverse relaxation times in liquid water, and is an extreme case. It is clear however, that electrostatic boundary conditions will affect the long time dynamics.
\begin{figure} [h]
\includegraphics[width=0.95\columnwidth]{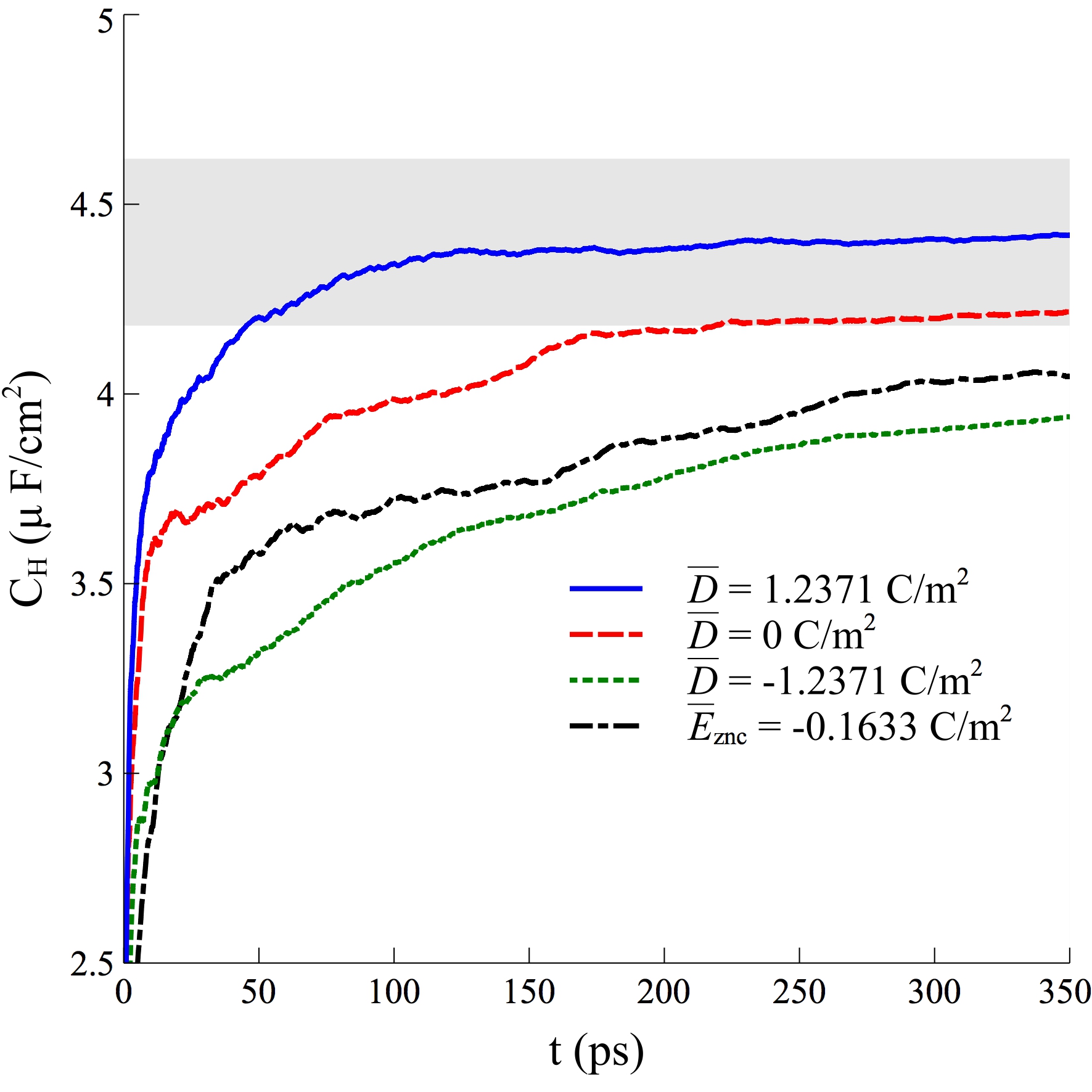}% Here is how to import EPS art
\caption{\label{taccum} The accumulating average of the calculated $C_{\textrm{H}}$  from Eq.~\ref{deltaP_D} at constant $\bar{D}$ for $\bar{D}=-1.2371$ C/m$^2$, $D=0$ C/m$^2$ and $\bar{D}=1.2371$ C/m$^2$ and from Eq.~\ref{polpzc} at constant $\bar{E}$ for $\bar{E}_{\textrm{znc}}=-0.1633$ C/m$^2$. A unit of the polarization quantum in $\bar{D}$ is $4\pi e/A=1.238$ C/m$^2$. The shaded region indicates a 5\% deviation from the target value. All calculations were carried out in the ECS geometry.}
\end{figure}

Fig.~\ref{taccum} compares the accumulating average of $C_{\textrm{H}}$ determined from Eq.~\ref{deltaP_D} at $\bar{D}=-1.2371$ C/m$^2$, $D=0$ C/m$^2$ and $\bar{D}=1.2371$ C/m$^2$ (recall that  the polarization quantum in $\bar{D}$ is $4\pi e/A=1.238$ C/m$^2$).  Also plotted is the accumulating average of  $C_{\textrm{H}}$ calculated from Eq.~\ref{polpzc} using the polarization at $\bar{E}_{\textrm{znc}}=-0.1633$ C/m$^2$. To keep the comparison fair, all simulations start from the same initial configuration extracted from an equilibrated system at zero plate charge. Therefore, this test will give an idea how fast the counterions from the electrolyte can rearrange to screen a finite insulator surface charge under different electric boundary conditions.

As the data of Fig.~\ref{taccum} show,  the calculation of capacitance at ZNC is more efficient with $\bar{D}=0$  compared to $\bar{E} = \bar{E}_{\textrm{znc}}$ electric boundary conditions. The data also confirm that the convergence time clearly depends on the value of $\bar{D}$.  The effect of $\bar{D}=0$ boundary conditions on the dynamics of ionic solutions has been studied by Caillol et al\cite{Caillol:1989jcpa, Caillol:1989jcpb}. The  $\bar{D}=0$ molecular dynamics was implemented by embedding the atomic system in an environment with dielectric constant $\epsilon^{\prime} =0$. As we have argued in Refs.~\citenum{Zhang:2015ms} and \citenum{Zhang:2016ho} this seemingly nonphysical model system is in fact a realization of a $\bar{D}=0$ system. Extension to finite $\bar{D}$ has become now possible with the development of the SSV constant $\bar{D}$ Hamiltonian (Eq.~\ref{uvdb}). This must remain for the moment the subject of future study.  However, while a theoretical analysis is still missing, it seems that it is feasible to optimize the magnitude of $\bar{D}$  for faster convergence of the capacitance calculation. This could well be an encouraging and desirable feature for DFT based MD modelling of charged insulator-electrolyte interface~\cite{Cheng:2014eh,Sulpizi:2016scirep,PfeifferLaplaud:2016cg}.

\section{Conclusion and outlook}

In this work, we applied the constant electric field methods of Stengel, Spaldin and Vanderbilt~\cite{Stengel:2009cd,Stengel:2009prb} in a calculation of the capacitance of a classical force field model of a charged insulator-electrolyte interface. Full periodic boundary conditions were applied confronting us with the classical version of multivalued polarization inherent in the modern theory of polarization\cite{King-Smith:1993prb,Resta:1994rmp,Resta:2007ch}. This could be resolved by consistent use of itinerant polarization already familiar from the molecular dynamics simulation of homogeneous ionic solutions\cite{Caillol:1989jcpa,Caillol:1989jcpb,Caillol:1994ho}. The key step was a comparison of the polarization of a ``safe'' electrolyte centred supercell locking in the ions and an open insulator centred supercell obtained by translation over half a MD box length. The latter effectively turns the cell inside out with boundaries bisecting the electrolyte allowing the ions to diffuse out.      

 Because of the finite width of the model insulator slab, the electric double layers formed at the insulator-electrolyte interface are not fully charge compensated. They bear a net charge which disappears only very slowly with the width of the insulator slab.  Periodic models of a size accessible to atomistic simulation are indeed a poor representation of charged insulator-electrolyte interfaces. This observation was rationalized using analytic expressions derived for a Stern-like continuum model. This model also suggested ways to recover the point of zero net charge using either the SSV constant $\mathbf{E}$ or constant $\mathbf{D}$ method each with a corresponding scheme for computing capacitance. Of these two schemes, we prefer the constant $\mathbf{D}$ method because capacitance is calculated from the change in polarization without having to compute the excess charge on the electrolyte side which in an electronic structure calculation may not be as straight forward as in a point charge model. 

The application of zero displacement field to aqueous electrolytes is not entirely new. As we have argued in previous work\cite{Zhang:2015ms, Zhang:2016ho}, $\epsilon'=0$ embedding method explored by Caillol and coworkers \cite{Caillol:1989jcpa,Caillol:1989jcpb,Caillol:1994ho} is equivalent to $\mathbf{D}=0$ electrostatic boundary conditions. This was also pointed out by Maggs in the context of the development of numerical schemes for solving the Poisson-Boltzmann equation using electric fields rather than electrostatic potentials\cite{Maggs:2004jcp}. However, the application of finite $\mathbf{D}$ goes one step further. This introduces an additional electrostatic control variable similar to the fixed charge on blocking electrodes attached to a system under open circuit\cite{Stengel:2009cd,Stengel:2009prb}. This opens up new ways of analyzing the electric response of interfaces as exploited by Stengel and Vanderbilt. This is also why the free charges (the ions) in our system were not treated as external charge but instead included in the polarization. A further advantage of making $\mathbf{D}$ available as an external control parameter is that charge relaxation can be accelerated by tuning the magnitude of the applied electric displacement. 

The motivation for the development presented in this paper is DFT based investigation of protonic double layers at oxide-electrolyte interfaces\cite{Cheng:2014eh,Cheng:2014angew}. The capacitance calculated for our model (4.4 $\mu$F/cm$^2$) is however much smaller than commonly reported for metal oxides, which is between 40 to 100 $\mu$F/cm$^2$ (see Refs.~\cite{Parez:2014dx,Cheng:2014eh}). The next step is therefore to revisit the DFT model of the TiO$_2$ interface of Ref.~\citenum{Cheng:2014eh}. In that work the authors  played it safe and used a symmetric (de)protonation scheme, which avoids the building up of an interior electric field in the slab. The capacitance was calculated to be between 30 and 40 $\mu$F/cm$^2$, which is of the right order of magnitude. The major gain of the constant field method presented here is that the slab can be charged at fixed composition. We should now be able to determine charging free energies which is not possible with the method of Ref.~\citenum{Cheng:2014eh}. Calculations are currently under way and will be reported in a forthcoming publication.

A further possibly artificial feature of our simple classical model is the remarkable linearity of the electric equations of state (Figs.~\ref{PDEeos}). This was in fact exploited in the calculation of capacitance at constant $\mathbf{D}$. Linearity enabled us to chose a convenient $\mathbf{D}$ even if this value was not at the point of zero net charge. There are several effects which in a realistic model could lead to non-linearity, such as dielectric saturation and coupling to stress\cite{Stengel:2009cd,Stengel:2009prb}.  In fact, the reason for the so called dielectric decrement at charged interfaces is still under debate\cite{Andelman:2011jcp}. Also, in DFTMD simulation the solid is free to move and  reorganization of the atomic structure of the insulator could also play a role. Finally we should point out that the continuum equations of state we have derived are  restricted to longitudinal polarization, i.e.~polarization perpendicular to dielectric discontinuities\cite{Matyushov:2014jcp}). For interfaces with an irregular geometry transverse polarization may have to be taken into account.  All these issues must also remain a subject of future investigation.

Despite the many open questions, we are optimistic that SSV constant field methods are a useful tool for quantitative DFT-based MD modelling of charged insulator-electrolyte interfaces. Finite size effects, while not eliminated, are less of an obstacle. The necessary times scales seem also in reach using the latest DFTMD methods\cite{Hutter:2013iea}. Prime candidates are redox active amorphous, hydrous or porous ceramic oxides\cite{Trasatti:1996cols} which are very hard to model using force fields. A priority for us is of course extending our studies of the catalytic activity of uncharged transition metal oxides\cite{Cheng:2014angew} to negatively charged surfaces at high pH. A further topic where these computational methods can contribute, is understanding the effect of an electrolyte on ferroelectric and polar surfaces\cite{Rappe:2009prl,Bristow:2012prb,Ismail-Beigi:2016ss}.  

\begin{acknowledgments}
Research fellowship (No.~ZH 477/1-1) provided by German Research
Foundation (DFG) for CZ is gratefully acknowledged. CZ and MS also
thank R.~Vuilleumier, R.~M.~Lynden-Bell and P. Wirnsberger
for helpful discussions. 
\end{acknowledgments}

%\nocite{*}
%\bibliography{abbr,EDL}% Produces the bibliography via BibTeX.

\end{document}